\documentclass[acmsmall]{acmart}
\usepackage[text=PREPRINT]{draftwatermark}
\usepackage{layouts}
\usepackage{listings}
\lstdefinelanguage
   [x64]{Assembler}     % add a "x64" dialect of Assembler
   [x86masm]{Assembler} % based on the "x86masm" dialect
   {morekeywords={CDQE,CQO,CMPSQ,CMPXCHG16B,JRCXZ,LODSQ,MOVSXD, %
                  POPFQ,PUSHFQ,SCASQ,STOSQ,IRETQ,RDTSCP,SWAPGS, %
                  rax,rdx,rcx,rbx,rsi,rdi,rsp,rbp, %
                  r8,r8d,r8w,r8b,r9,r9d,r9w,r9b, %
                  r10,r10d,r10w,r10b,r11,r11d,r11w,r11b, %
                  r12,r12d,r12w,r12b,r13,r13d,r13w,r13b, %
                  r14,r14d,r14w,r14b,r15,r15d,r15w,r15b}} % etc.

\lstdefinestyle{C}{
    language=C,
    keywordstyle=\color{blue},
    stringstyle=\color{red},
    commentstyle=\color{olive},
}

\lstdefinestyle{tinyC}{
    language=C,
    basicstyle={\tiny\ttfamily},
    keywordstyle=\color{blue},
    stringstyle=\color{red},
    commentstyle=\color{olive},
}

\lstdefinestyle{asm}{
    language=[x64]Assembler,
    keywordstyle=\color{blue},
    stringstyle=\color{red},
    commentstyle=\color{olive},
}

\usepackage{xcolor}
\usepackage{pgf}
\usepackage{import}
\usepackage{tikz}
\usetikzlibrary{calc}
\usetikzlibrary{trees}
\usetikzlibrary{shapes.geometric}
\usepackage{pifont}
\usetikzlibrary{shapes}
\usetikzlibrary{decorations.pathreplacing,calligraphy}
\usetikzlibrary{shadings}

\DeclareRobustCommand*\circledColorSmall[2]{\tikz[baseline=(char.base)]{
    \node[shape=circle,fill=#2,draw=#2,inner sep=0pt] (char) {\textcolor{white}{\footnotesize\textbf{#1}}};}} 

\usepackage{amsmath}
\usepackage{mathtools}
\usepackage{float}
\usepackage{xcolor}
\usepackage{semantic}
\usepackage{algpseudocodex}
\usepackage{dblfloatfix}

\usepackage[frozencache,cachedir=.]{minted}
\setminted[C]{
    frame=lines,
    framesep=5pt,
    fontfamily=tt,
    fontsize=\footnotesize,
    escapeinside=@@,
    breaklines=true,
    stripnl=false,
}

\setminted[gas]{
    frame=lines,
    framesep=5pt,
    fontfamily=tt,
    fontsize=\footnotesize,
    escapeinside='',
    breaklines=true,
    stripnl=false,
}

\algnewcommand\algorithmicforeach{\textbf{for each:}}
\algnewcommand\ForEach{\item[ \algorithmicforeach]}

\definecolor{boxbg}{rgb}{0.9,0.9,0.9}
\definecolor{dusty_red}{HTML}{C44E52}
\definecolor{french_blue}{HTML}{4C72B0}
\definecolor{algae}{HTML}{55A868}
\definecolor{terracota}{HTML}{DD8452}
\definecolor{p_green}{HTML}{66a61e}
\definecolor{p_orange}{HTML}{d95f02}
\colorlet{comment}{olive}

\usepackage{pgfplots}
\usepgfplotslibrary{colorbrewer}
\usepgfplotslibrary{groupplots}
\usetikzlibrary{plotmarks}
\usetikzlibrary{positioning, arrows.meta}
\usetikzlibrary{decorations.text}
 \usetikzlibrary{patterns}

\pgfplotsset{
  compat=1.17,
  every axis/.append style={
    axis line style={-latex},
    xlabel style={font=\small},
    ylabel style={font=\small},
    ymajorgrids,
    grid style={densely dotted},
    tick label style={font=\footnotesize},
    legend style={font=\footnotesize},
    title style={align=center},
    cycle list/Dark2,
    cycle multiindex* list={
            mark list*\nextlist
            Dark2\nextlist
    },
  },
}

\AtBeginDocument{%
  \providecommand\BibTeX{{%
    \normalfont B\kern-0.5em{\scshape i\kern-0.25em b}\kern-0.8em\TeX}}}

\setcopyright{rightsretained}
\copyrightyear{2024}
\acmYear{2024}
\acmDOI{10.1145/3701993}

\acmJournal{TACO}
\begin{document}

\title{Iterating Pointers: Enabling Static Analysis for Loop-based Pointers}

\author{Andrea Lepori}
%\email{webmaster@marysville-ohio.com}
\affiliation{%
  \institution{ETH Zurich}
  %\streetaddress{P.O. Box 1212}
  %\city{Zurich}
  \country{Switzerland}
}
\email{andrea.lepori@inf.ethz.ch}

\author{Alexandru Calotoiu}
\affiliation{%
  \institution{ETH Zurich}
  %\streetaddress{P.O. Box 1212}
  %\city{Zurich}
  \country{Switzerland}
}
\email{acalotoiu@inf.ethz.ch}

\author{Torsten Hoefler}
\affiliation{%
  \institution{ETH Zurich}
  %\streetaddress{P.O. Box 1212}
  %\city{Zurich}
  \country{Switzerland}
}
\email{torsten.hoefler@inf.ethz.ch}

\renewcommand{\shortauthors}{Lepori et al.}

\begin{abstract}
Pointers are an integral part of C and other programming languages. They enable substantial flexibility from the programmer's standpoint, allowing the user fine, unmediated control over data access patterns. However, accesses done through pointers are often hard to track, and challenging to understand for optimizers, compilers, and sometimes, even for the developers themselves because of the direct memory access they provide. We alleviate this problem by exposing additional information to analyzers and compilers. By separating the concept of a pointer into a data container and an offset, we can optimize C programs beyond what other state-of-the-art approaches are capable of, in some cases even enabling auto-parallelization. Using this process, we are able to successfully analyze and optimize code from OpenSSL, the Mantevo benchmark suite, and the Lempel–Ziv–Oberhumer compression algorithm. We provide the only automatic approach able to find all parallelization opportunities in the HPCCG benchmark from the Mantevo suite the developers identified and even outperform the reference implementation by up to 18\%, as well as speed up the PBKDF2 algorithm implementation from OpenSSL by up to 11x. 
\end{abstract}

\begin{CCSXML}
<ccs2012>
   <concept>
       <concept_id>10011007.10011006.10011041.10011047</concept_id>
       <concept_desc>Software and its engineering~Source code generation</concept_desc>
       <concept_significance>500</concept_significance>
       </concept>
   <concept>
       <concept_id>10010147.10010169</concept_id>
       <concept_desc>Computing methodologies~Parallel computing methodologies</concept_desc>
       <concept_significance>500</concept_significance>
       </concept>
 </ccs2012>
\end{CCSXML}

\ccsdesc[500]{Software and its engineering~Source code generation}
\ccsdesc[500]{Computing methodologies~Parallel computing methodologies}

\keywords{pointer analysis, automatic parallelization}


\maketitle

\section{Introduction}
\label{sec:intro}
\let\thefootnote\relax\footnotetext{New Paper, Not an Extension of a Conference Paper}

C is a widely used programming language, one that is heavily relied upon in many codebases. In 2022 on GitHub C and C++ accounted together for 4.6M repositories~\cite{ghapi}, this being a 23.5\% growth compared to the previous year~\cite{ghlang}. According to the TIOBE index~\cite{tiobe}, C is the second most popular programming language.

The C language was developed in the 1970s at about the same time as the first microprocessor~\cite{10.1145/234286.1057834, intel4004}. Parallel computing only became widespread decades later, driven by the end of Dennard scaling~\cite{bell1994scalable, 4785860}. Hardware efforts to continue growth following Moore's Law past the power wall lead to a renewed focus on parallel computing as a way to continue improving performance~\cite{gagliardi2019international, asanovic2006landscape}.
C was not designed with parallel programming in mind as its inception was much earlier than the explosion of parallel computing. Libraries provide support for parallel paradigms - such as OpenMP~\cite{openmp} or MPI~\cite{mpi40} - but add a difficult-to-manage layer of complexity. The flexibility and low-level nature of C code make debugging challenging, even more so when the code is executed in parallel~\cite{10.1145/115372.115324, 1137778}. This makes parallelizing code a long and complex process.

Frameworks such as Polly~\cite{polly} and Pluto~\cite{pluto} promise automatic parallelism  --- these are however limited to static control parts (SCoPs). The Intel compiler offers the \texttt {icc parallel} mode that can identify and auto-parallelize several predefined programming patterns. Data-centric methods~\cite{c2dace} try to push the boundary even further by leveraging complex dataflow analysis techniques to better understand data movement and dependencies. On modern systems, data movement is the most expensive operation in most programs, concerning both time and energy consumption~\cite{10.48550}, and tends to be the biggest bottleneck in computations~\cite{9530719}. Many frameworks and compilers such as HPVM~\cite{10.1145/3178487.3178493}, Halide~\cite{10.1145/3150211}, Jax~\cite{47008}, and DaCe~\cite{dace} leverage dataflow analysis extensively in their pipeline.

The major obstacle in the static analysis of data accesses is unfortunately the use of pointers. Though they are an integral part of the C language, they often become a barrier on the road to performance. This can be seen in the two examples in Figure~\ref{intro_code}. On the right side of the figure we can see that the inner loop only has independent data accesses, and modern compilers are able to detect that. But no compiler we tested was able to detect that the outer loop has no data dependency. Every access done with \texttt{p[k]} is different because of the pointer movement in the outer loop. Current analysis methods do not consider this type of access as it would require complex and expensive pointer tracking, and even then it would require a correlation between the pointer movement and loop iterations.

\begin{figure}[h] 
\begin{minipage}[t]{.48\textwidth}
Accessing the same pointer at a different index does not guarantee that different data is accessed --- pointers can move during execution.
\end{minipage}
\hfill
\begin{minipage}[t]{.5\textwidth}
Accessing the same pointer at the same location with access to different data. Creation of false positives in data dependency analysis.
\end{minipage}

\begin{minipage}[t]{.48\textwidth}
\begin{minted}{C}
p[i+1] = 0; // array-style access
p++;        // pointer moves
p[i] = 42;  // array-style access to same memory
\end{minted}
\end{minipage}
\hfill
\begin{minipage}[t]{.5\textwidth}
\begin{minted}{C}
for (int i=0; i<cplen*n; i+=cplen) {
  for (int k=0; k<cplen; k++) {
    p[k] = p[k] ^ data[k]; // no loop dependence
  }
  p += cplen; // outer iteration on different data
}
\end{minted}
\tiny{Snippet of the \texttt{PBKDF2} implementation from OpenSSL}
\end{minipage}

\caption{Two example codes where we can have false positives and negatives with pointer accesses}
\label{intro_code}
\end{figure}

Our solution aims to address this challenge by tracking data containers (as opposed to pointers) across the entire program. We then split pointers into the data container and an integer index to it. This effectively makes data accesses explicit while preserving execution semantics, enabling better analyzability of the code. A state-of-the-art data-centric compiler can then attempt to parallelize and improve the performance of the code. The implementation of our methods is explained in Section~\ref{implementation}. Afterwards in Section~\ref{limitations} we discuss the limitations and motivation behind them. In Section~\ref{specific_transformations} we will discuss possible solutions to mitigate the limitations showing some specific transformations handling special cases.

\break

\textit{Contributions}
\begin{itemize}
    \item We introduce a novel method to statically split pointers into a data container and an index to it, creating explicit data accesses to aid further analysis.
    \item We generate a parallel version of the Mantevo HPCCG benchmark automatically finding all parallelization opportunities the benchmark developers envisioned manually. Our automatic workflow outperforms the manually tuned version by up to 6\%.
    \item We automatically parallelize the previously serial PBKDF2 algorithm implemented in OpenSSL and obtain up to an 11x speedup.
\end{itemize}

\section{Pointer disaggregation} \label{implementation}
We propose a static method to improve the analyzability of pointers that are used as iterators over data containers. We first provide a high level overview explaining our method, we follow with some notes toward a proof and the description of the technical implementation. Then we compare our method with previous work. We also provide some examples and results to show the value of our transformation.

\subsection{High level overview}

\begin{figure}[h]
    \begin{minipage}{0.49\linewidth}
    \hrule
    \vspace{0.3em}
    \textit{Original code with pointer movements\\}
    \vspace{-1em}
    \begin{minted}[fontsize=\small]{C}
@\phantom@
p = a;    @\tikzmark{ptr_a}@






p += i;   @\tikzmark{ptr_a_2}@


x = p[k];
@\tikzmark{annotate}@
    \end{minted}
    \begin{tikzpicture}[remember picture, overlay, comment]
        \tikzset{shift={(pic cs:ptr_a)}};
        \tikzset{shift={(0, -0.06)}};
        \draw (0, 4pt) node[anchor=east] {\texttt{a}};
        \draw[step=8pt, opacity=0.6] (0,0) grid (120pt,8pt);
        
        \draw[-latex] (5pt, 14pt) -- (4pt, 9pt);
        \draw (8pt, 12pt) node[anchor=south] {\texttt{p}};
        
        \draw[pen colour={comment}, decorate,decoration={calligraphic brace, mirror}] (4pt, -2pt) -- node[below,align=center]{\texttt{i} elements} +(56pt, 0);
    \end{tikzpicture}
    \begin{tikzpicture}[remember picture, overlay, comment]
        \tikzset{shift={(pic cs:ptr_a_2)}};
        \tikzset{shift={(0, -0.06)}};
        \draw (0, 4pt) node[anchor=east] {\texttt{a}};
        \draw[step=8pt, opacity=0.6] (0,0) grid (120pt,8pt);
        
        \draw[-latex] (8pt, 21pt) -- ++(52pt, 0) -- ++(0, -12pt);
        \draw (4pt, 14pt) node[anchor=south] {\texttt{p}};
    
        \draw[-latex] (100pt, 14pt) -- +(0, -5pt);
        \draw (100pt, 14pt) node[anchor=south, xshift=1em] {\texttt{p[k] = x}};
    
        \draw[pen colour={comment}, decorate,decoration={calligraphic brace, mirror}] (60pt, -2pt) -- node[below,align=center]{\texttt{k} elements} +(40pt, 0);
    \end{tikzpicture}
    \begin{tikzpicture}[remember picture, overlay, comment]
        \tikzset{shift={(pic cs:annotate)}};
        \draw (60pt, 4pt) node (container) {container};
        \draw[-latex] (container.west) -- (24pt, 4pt) -- (20pt, 8pt);
        
        \draw (190pt, 4pt) node (container2) {container};
        \draw[-latex] (container2.east) -- (216pt, 4pt) -- (220pt, 8pt);
        
        \draw (60pt, 25pt) node (offset) {offset};
        \draw[-latex] (offset.west) -- (34pt, 25pt) -- (30pt, 18pt);
        \draw[-latex] (34pt, 25pt) -- (24pt, 25pt) -- (20pt, 18pt);

        \draw (200pt, 25pt) node (offset2) {offset};
        \draw[-latex] (offset2.east) -- (232pt, 25pt) -- (240pt, 18pt);
        \draw[-latex] (232pt, 25pt) -- (262pt, 25pt) -- (270pt, 18pt);
    \end{tikzpicture}
    \end{minipage}
    \hfill
    \begin{minipage}{0.49\linewidth}
    \hrule
    \vspace{0.3em}
    \textit{Transformed code with \textit{adjunct} instead of pointer movements}
    \vspace{-1em}
    \begin{minted}[fontsize=\small]{C}
@\phantom@
p = a;
p_adj = 0;    @\tikzmark{adjunct_a}@





p_adj += i;   @\tikzmark{adjunct_a_2}@


x = p[p_adj + k];
@\phantom@
    \end{minted}
    \begin{tikzpicture}[remember picture, overlay, comment]
        \tikzset{shift={(pic cs:adjunct_a)}};
        \tikzset{shift={(0, -0.06)}};
        \draw (0, 4pt) node[anchor=east] {\texttt{a}};
        \draw[step=8pt, opacity=0.6] (0,0) grid (120pt,8pt);
        
        \draw[-latex] (4pt, 14pt) -- +(0, -5pt);
        \draw (4pt, 14pt) node[anchor=south] {\texttt{p}};
    
        \draw[-latex] (38pt, 21pt) -- ++(-30pt, 0) -- (6pt, 9pt);
        \draw (60pt, 14pt) node[anchor=south] {\texttt{p[p\_adj]}};
        
        \draw[pen colour={comment}, decorate,decoration={calligraphic brace, mirror}] (4pt, -2pt) -- node[below,align=center]{\texttt{i} elements} +(56pt, 0);
    \end{tikzpicture}
    \begin{tikzpicture}[remember picture, overlay, comment]
        \tikzset{shift={(pic cs:adjunct_a_2)}};
        \tikzset{shift={(0, -0.06)}};
        \draw (0, 4pt) node[anchor=east] {\texttt{a}};
        \draw[step=8pt, opacity=0.6] (0,0) grid (120pt,8pt);
        
        \draw[-latex] (4pt, 14pt) -- +(0, -5pt);
        \draw (4pt, 14pt) node[anchor=south] {\texttt{p}};
        
        \draw[-latex] (60pt, 14pt) -- +(0, -5pt);
        \draw (60pt, 14pt) node[anchor=south] {\texttt{p[p\_adj]}};
    
        \draw[-latex] (100pt, 28pt) -- (100pt, 9pt);
        \draw (80pt, 28pt) node[anchor=south, xshift=0.5em] {\texttt{p[p\_adj + k] = x}};
        
        \draw[pen colour={comment}, decorate,decoration={calligraphic brace, mirror}] (4pt, -2pt) -- node[below,align=center]{\texttt{p\_adj} elements} +(56pt, 0);
    
        \draw[pen colour={comment}, decorate,decoration={calligraphic brace, mirror}] (60pt, -2pt) -- node[below,align=center,xshift=1.5em]{\texttt{k} elements} +(40pt, 0);
    \end{tikzpicture}
    \end{minipage}
    \vspace{-1em}

    \caption{Representation of data access patterns with pointer movement (left) and static accesses with \textit{adjunct} (right). Annotated with the accessed memory locations and pointer values. In the case of pointer movements the pointer is both the container and the offset while with the \textit{adjunct} the pointer is only the container.}
    \label{ptr_adjunct}
\end{figure}

Pointers are used both as handles for data containers and to iterate over them. By splitting these two semantically different use cases we intend to improve element-sensitive analysis of pointer accesses. For each pointer, we create an \textit{adjunct} \texttt{int} variable that represents the offset from the start of the data container. Note that the type of the \textit{adjunct} should be big enough to address every element plus one in the connected data-container. The goal of the \textit{adjunct} is to allow iteration over the data container without modifying the handle used to access said data container. We need the \textit{adjunct} type to be plus one larger than the size of the data container as with loops containing \texttt{*p++} we could have the need to get the address after the last element.

Figure~\ref{ptr_adjunct} shows a high-level example of how the \textit{adjunct} is used. We can see from the figure that the value of \texttt{x} after both executions is the same. For general code, it is important to note that \texttt{p[i]} is semantically equivalent to \texttt{*(p + i)} as per C99 standard~\cite{ISO:C99}. In Figure~\ref{adjunct_example} we present a minimal example to show the value of introducing the \textit{adjunct} variable. We provide the resulting assembly to show that when using the \textit{adjunct}, the compiler is able to correctly identify that $p_{adj}$ is only used as an iterator similar to $i$ and merges the two, resulting in more compact, efficient code. 

\begin{figure*}[h]
    \begin{minipage}{.30\textwidth}
        \textit{C representation}
        \begin{minted}{C}
for (int i=0; i<n; i++) {
  p[0] = i + 1;
  p++;
}
@\phantom@
@\phantom@
        \end{minted}
    \end{minipage}
    \hfill
    \begin{minipage}{.68\textwidth}
        \textit{ASM representation}
        \begin{minted}{gas}
.L4:
  add     eax, 1     ; i++
  add     rdx, 4     ; p++
  mov     DWORD PTR [rdx-4], eax
  cmp     eax, r14d  ; i < n
  jne     .L4
        \end{minted}
    \end{minipage}

\vspace{1cm}
    
    \begin{minipage}{.30\textwidth}
        \begin{minted}{C}
for (int i=0; i<n; i++) {
  p[p_adj + 0] = i + 1;
  p_adj++;
}
@\phantom@
        \end{minted}
    \end{minipage}
    \hfill
    \begin{minipage}{.68\textwidth}
    {
        \renewcommand\fcolorbox[4][]{\textcolor{cyan}{\strut#4}}
        \begin{minted}{gas}
.L4:
  mov     DWORD PTR [r12-4+rax*4], eax
  add     rax, 1     ; i++
  cmp     rdx, rax   ; i < n
  jne     .L4
        \end{minted}
    }
    \end{minipage}
    
    \begin{tikzpicture}[overlay]
        \draw[ultra thick, -latex, olive] (3.7, 3.05) -- +(0, -0.9);
        \draw[olive] (3.6, 2.7) node[anchor=east, align=left] {
            \textbf{No pointer increment}\\
            \textbf{$\Rightarrow$ one less instruction}
        };
        %\draw[olive] (5.8, 2.55) node[anchor=west, align=left] {
        %    \textbf{One less register}
        %};
        
        \draw[ultra thick, -latex, olive] (-3, 3.05) -- +(0, -0.9);
        \draw[olive] (-3, 2.7) node[anchor=east, align=left] {
            \textbf{\textit{adjunct} transformation}
        };

        \draw[olive] (3, 4.45) node[anchor=west, align=left] {
            loop iterator $\rightarrow$ \texttt{eax}\\
            pointer iterator $\rightarrow$ \texttt{rdx}\\
            $n$ value $\rightarrow$ \texttt{r14d}
        };
        
        \draw[olive] (3, 1.2) node[anchor=west, align=left] {
            loop iterator $\rightarrow$ \texttt{rax}\\
            pointer iterator $\rotatebox[origin=c]{270}{$\Lsh$}$\\
            container base address $\rightarrow$ \texttt{r12}\\
            $n$ value $\rightarrow$ \texttt{rdx}
        };
    \end{tikzpicture}
    \caption{Difference in the assembly representation (compiled with \texttt{gcc} 12.2 and \texttt{-O2}) between a code using pointer movement and one with static access after the \textit{adjunct} is introduced.}
    \label{adjunct_example}
\end{figure*}

The \textit{adjunct} transformation enhances code analyzability and exposes additional parallelization opportunities. Such transformed pointers are equivalent to arrays, in that the symbol represents a data container that is statically known. After the transformation the \textit{adjunct} variable dictates how the pointer is accessed. Because it is an \texttt{int} variable there are more compiler analysis methods available to track the value. Inside loops a compiler could apply Loop Strength-Reduction (LSR) to simplify addressing computations. Note that this can only be done after the \textit{adjunct} transformation is applied as LSR only acts on integer values and not addresses (pointers). If instead the \textit{adjunct} variable is never used - if a pointer is never moved for example - then the compiler will be able to optimize away the \textit{adjunct} with constant propagation resulting in no performance losses. A summary of the transformation is described in Figure~\ref{tab:transform}. To note is that the expression \texttt{p = q $\diamond$ x} is a combination of \texttt{p = q; q = q $\diamond$ x}. Because of this fact, we do not detail this case in the following sections as it follows from the other statements.

The \textit{adjunct} type is deliberately a \textit{signed} integer to replicate the behavior of actual pointer movement. This is because of the non-wrapping behaviour when the \textit{adjunct} becomes negative. It is worth noting that accessing a pointer with a negative offset might be considered valid behavior under certain circumstances. For instance, if a programmer is confident that the memory is initialized within the same program, such as when an opaque library provides a pointer within an array, and the programmer is aware of the underlying structure.

\begin{figure}[h]
\begin{tabular}{ll}
\toprule
Original code & Modified code\\
\midrule
\texttt{int* p;} & \texttt{int* p;} \\
& \texttt{int p\_adj = 0;} \\[5pt]
\texttt{*p} & \texttt{p[p\_adj]} \\[5pt]
\texttt{p[x]} & \texttt{p[x + p\_adj]} \\[5pt]
\texttt{p = p $\diamond$ x;} & \\
or \texttt{p $\diamond$= x;} & \texttt{p\_adj = p\_adj $\diamond$ x;} \\[5pt]
\texttt{p = q;} & \texttt{p = q;} \\
& \texttt{p\_adj = q\_adj;} \\[5pt]
\texttt{p = q $\diamond$ x;} & \texttt{p = q;} \\
& \texttt{p\_adj = q\_adj $\diamond$ x;} \\
\bottomrule
\end{tabular}
\quad
\begin{tabular}{ll}
\toprule
Original code & Modified code\\
\midrule
\texttt{f(p);} & \texttt{f(p + p\_adj);} \\[5pt]
\texttt{void f(int* p) \{} & \texttt{void f(int* p) \{} \\
& \texttt{  int p\_adj = 0;} \\
\texttt{    // body} & \texttt{  // body} \\
\texttt{    int k = *p;} & \texttt{    int k = p[p\_adj];} \\
\texttt{\}} & \texttt{\}} \\
\bottomrule
\end{tabular}
 \caption{Transformation summary to improve pointer analyzability. Both \texttt{p} and \texttt{q} are \texttt{int} pointers (\texttt{int* p, q}), \texttt{x} is an \texttt{int} expression and $\diamond$ is any integer binary operator. Note that the data-type \texttt{int} is only for explanation purposes and can be substituted with any other type. The \textit{adjunct}-type \texttt{int} can be interchanged with any integer type that fits the size of the data-container.}
    \label{tab:transform}
\end{figure}

Function arguments and calls undergo minimal modifications. When a function is called, it is treated as a dereference, and we pass the pointer with the \textit{adjunct} offset applied. The function argument declarations remain unchanged, meaning that the \textit{adjunct} information is lost after the function call. This is generally not an issue, as subroutines often do not require the additional information. Moreover, this problem is mitigated by the compiler's automatic inlining or by manually annotating functions to force inlining, thereby eliminating the function call.

\subsection{Notes toward a proof}
We provide a proof sketch that the transformation showed in Figure~\ref{tab:transform} does not modify the accesses done with pointers. First for initialization of a pointer let $p$ be a pointer to some data. Then let $p_{adj}$ be a new unique integer variable not already present in the program and initialized to zero when the memory of the pointer is first initialized. I.e. we have $p = malloc(...); p_{adj} = 0$ where $p_{adj}$ was not previously present in the variable definitions. This is a new variable definition without any name conflict hence it will not influence other parts of the program.

Let $S$ be an arbitrary $n$ length sequence of statements s.t.
$$S = \left (p := p + x_i | x_i \in \mathbb{Z}, 0 \leq i < n\right )$$
Then we define the syntax $S; p$ by using the sequential operator $;$ with a variable name at the end as the value of $p$ after execution of $S$. By our definition we have that:
$$(S; p) = p + \sum_{i=0}^n x_i$$

Let $S'$ be an $n$ length sequence of statements (note that $x_i$ are the same values as in $S$ and in the same order).
$$S' = \left (p_{adj} := p_{adj} + x_i | x_i \in \mathbb{Z}, 0 \leq i < n\right )$$
As before by definition we have that
$$(S'; p_{adj}) = p_{adj} + \sum_{i=0}^n x_i$$

We can see that
$$(S; p) = p + \sum_{i=0}^n x_i = p + p_{adj} + \sum_{i=0}^n x_i - p_{adj} = p + (S'; p_{adj}) - p_{adj}$$

By using the sequential operator $;$ we define a program sequence that defines the pointer $p$, integer $p_{adj}$ and executes all statements in $S$. This represents an arbitrary program from the definition to the access of a pointer $p$ with arbitrary pointer movements in between.
\begin{align*}
(p := malloc(...); (p_{adj} := 0; (S; p))) &= (p := malloc(...); (p_{adj} := 0; (p + (S'; p_{adj}) - p_{adj}))) \\
&= (p := malloc(...); (p_{adj} := 0; (p + (S'; p_{adj}))))
\end{align*}

We note that $(S'; p) = p$ and $(S; p_{adj}) = p_{adj}$ as the statements does not affect the evaluated variable. Then we have that if we substitute any execution $p = p + x_i$ with $p_{adj} = p_{adj} + x_i$ i.e. executing $S'$ instead of $S$ then
\begin{align*}
(p := malloc(...); (p_{adj} := 0; (S; p))) &= (p := malloc(...); (p_{adj} := 0; (p + (S'; p_{adj})))) \\
&= (p := malloc(...); (p_{adj} := 0; (S'; p) + (S'; p_{adj}))) \\
&= (p := malloc(...); (p_{adj} := 0; (S'; p + p_{adj})))
\end{align*}

This means that by substituting $S$ with $S'$ and accessing every pointer using $p + p_{adj}$ as their address we access the same memory location. Hence the program will be semantically equivalent. $S$ and $S'$ only contain statements referring to $p$ or $p_{adj}$ for simplicity of the proof. Any other statement could be interleaved including control flow statements. By getting all possible execution paths and removing all statements that don't act directly on $p$ then we have the same case as $S$ that can be transformed in $S'$.

The proof sketch for the correctness of the pointer assignment transformation can be found in Appendix~\ref{appendix_assignment}.

\subsection{Technical implementation}
We will now discuss the algorithm that governs our approach.
The algorithms are applied to the AST (abstract syntax tree) to produce the modified code including the \textit{adjunct} transformation. It is important to note that the whole transformation is source-to-source.
The implementation leverages \texttt{libclang} to parse the source code. We keep a $adjunctMapping$ map that maps pointers to its \textit{adjunct}. As previously mentioned the type of the \textit{adjunct} variable should be an integer type big enough to address each element of the data-container. We mostly use \texttt{int} as it is big enough for most practical purposes but it could be easily interchanged with bigger or smaller integer types.

For every variable declaration, we check if it is a pointer type. In that case, we append an additional variable declaration for the \textit{adjunct} variable with type \texttt{int} and initialize it to 0. We keep a mapping between the pointer variables and their \textit{adjunct}. To avoid name collisions for the \textit{adjunct} variable name a simple counter and checks are used.

\hfill
\noindent
\hrule
\vspace{0.1cm}
\begin{algorithmic}
\ForAll {\text{variable declaration of the form } $type \text{ } name = init$}
    \If {$type $ is a pointer type}
        \State $adjunctName \gets getUniqueName(name +  ``\_adj\text{''})$
        \State $adjunctMapping[name] \gets adjunctName$
        \State $adjunctName \text{ init to 0}$
    \EndIf
\EndFor
\end{algorithmic}
\vspace{0.1cm}
\hrule
\vspace{0.1cm}

\hfill

We analyze every array subscript expression (\texttt{a[i]}, where \texttt{a} is the array expression and \texttt{i} is the index expression) and verify if the array expression is in the mapping. In case we have a match we modify the index expression to sum the additional \textit{adjunct} variable. Note that we can handle unary expressions of the type \texttt{*p} easily by transforming them to \texttt{p[0]} with a simple transformation (which does not change the semantics as per C standard \texttt{E1[E2]} is equivalent to \texttt{*(E1 + E2)}).

\break

\noindent
\hrule
\vspace{0.1cm}

\begin{algorithmic}
\ForAll {\text{array subscript expression of the form } $array[index]$}
    \If {$array \in adjunctMapping$}
        \State $adjunct \gets adjunctMapping[array]$
        \State $index = index + adjunct$
    \EndIf
\EndFor
\end{algorithmic}
\vspace{0.1cm}
\hrule
\vspace{0.1cm}

\hfill

Finally, we examine every assignment. Here we have two cases: \texttt{p = p $\diamond$ x} or \texttt{p = q}. For both, the left expression must be in the mapping. In the first case (\texttt{p = p $\diamond$ x}) the right side must be a binary operation for which the first operator is the same as the left side of the assignment (\texttt{p = q $\diamond$ x}, \texttt{p} and \texttt{q} must be the same pointer). We then substitute the pointer \texttt{p} with its \textit{adjunct} (\texttt{p\_adj = p\_adj $\diamond$ x}). In the second case (\texttt{p = q}) we check that the right side is also a pointer in the mapping. Then we append an additional statement that overwrites the \textit{adjunct} of the left side with the \textit{adjunct} of the right (\texttt{p\_adj = q\_adj}).

\noindent
\vspace{0.1cm}
\hrule
\vspace{0.1cm}
\begin{algorithmic}
\ForAll {\text{assignment of the form } $lhs = rhs$}
    \If {$lhs \in adjunctMapping$}
        \If {$rhs \text{ is a pointer } \wedge rhs \in adjunctMapping$}
            \State $lhsAdjunct \gets adjunctMapping[lhs]$
            \State $rhsAdjunct \gets adjunctMapping[rhs]$
            \State $lhsAdjunct = rhsAdjunct$
        \ElsIf {$rhs == ptr \diamond expr$}
            \If {$lhs == ptr$}
                \State $lhsAdjunct \gets adjunctMapping[lhs]$
                \State $lhs = lhsAdjunct$
                \State $rhs = lhsAdjunct \diamond expr$
            \EndIf
        \EndIf
    \EndIf
\EndFor
\end{algorithmic}
\vspace{0.1cm}
\hrule
\vspace{0.1cm}

\subsection{Previous approaches}
A similar approach was already done specifically for digital signal processing (DSP) applications~\cite{185487}. Their approach could also be extended to general programs and modifies the source code to remove pointer movement. Our biggest innovation is the use of the \textit{adjunct} that enables run-time determined pointer movements. Previous methods only handled static or compile-time decidable pointer increments.

For example the expression:
\begin{minted}{C}
if (exp) {
    x = ptr++;
} else {
    y = ptr + 2;
}
\end{minted}
cannot be handled by previous approaches as it cannot decide at compile time if the pointer must be incremented by 1 or 2. With our \textit{adjunct} approach it can be handled as the same branch just by modifying the \textit{adjunct} instead of the pointer.

\begin{figure}[ht]
    \noindent\begin{minipage}{0.49\linewidth}
    \begin{minted}{C}
int* arr = malloc(sizeof(int) * 10);
// random init array

int* p = arr;

int* q = arr;


if (argc == 42) {
    p += 5;
} else {
    q += 5;
}

for (int i=0; i<5; i++) {
    // GCC adds runtime vectorization checks
    p[i] += q[i]; @\tikzmark{non_vec}@
}
    \end{minted}
    \begin{tikzpicture}[remember picture, overlay, olive]
        \tikzset{shift={(pic cs:non_vec)}};
        \tikzset{shift={(0, 0.06)}};
        \draw[-latex, thick] (1.1, 0.5) -- node[
                                pos=0.7, anchor=east, align=right] {\footnotesize Unrolled loop\\[-5pt] \footnotesize no vectorization} (1.1, 1.85);
        \draw[-latex, thick] (2.6, 0.5) .. controls (2.6, 1.8) .. node[
                                pos=0.2, anchor=east, align=right] {\footnotesize Unrolled\\[-5pt] \footnotesize vectorized \\[-5pt] \footnotesize loop} (2.8, 2.3);
        \node[anchor=north west] at (0.5, 5.5) {\includegraphics[height=4.5cm]{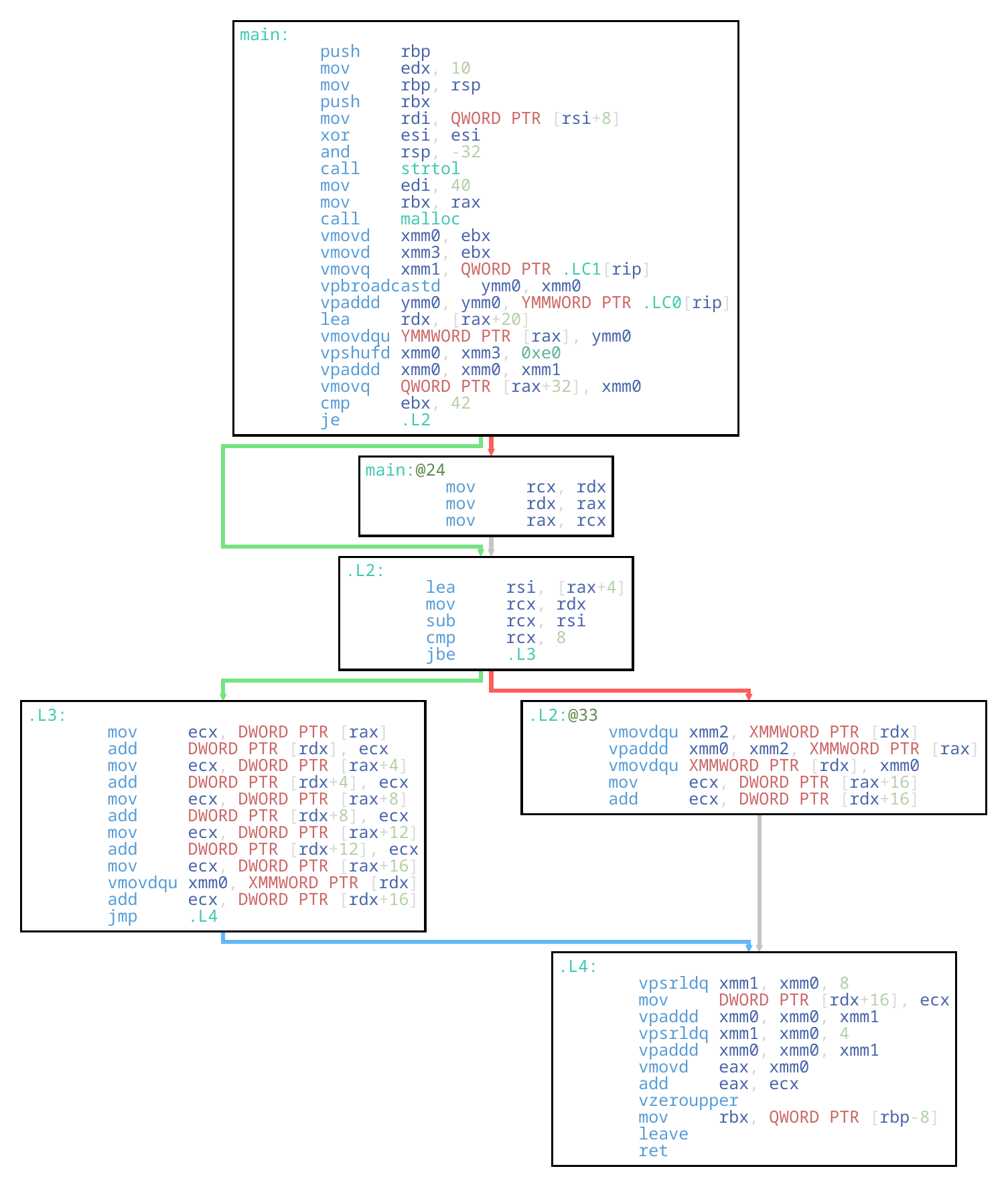}};
    \end{tikzpicture}
    \end{minipage}
    \hfill
    \begin{minipage}{0.49\linewidth}
    \begin{minted}{C}
int* arr = malloc(sizeof(int) * 10);
// random init array

int* p = arr;
int p_adj = 0;
int* q = arr;
int q_adj = 0;

if (argc == 42) {
    p_adj += 5;
} else {
    q_adj += 5;
}

for (int i=0; i<5; i++) {
    // GCC statically decided vectorization
    p[i + p_adj] += q[i + q_adj]; @\tikzmark{vec}@
}
    \end{minted}
    \begin{tikzpicture}[remember picture, overlay, olive]
        \tikzset{shift={(pic cs:vec)}};
        \tikzset{shift={(0.3, 0.06)}};
        \draw[-latex, thick] (-1.4, 0.5) .. controls (-1.5, 2.2) and (-1, 2.5) .. node[
                                pos=0.2, anchor=east, align=right] {\footnotesize Unrolled\\[-5pt] \footnotesize vectorized \\[-5pt] \footnotesize loop} (-0.3, 2.5);
        \node[anchor=north west] at (-1, 5.5) {\includegraphics[height=4.5cm]{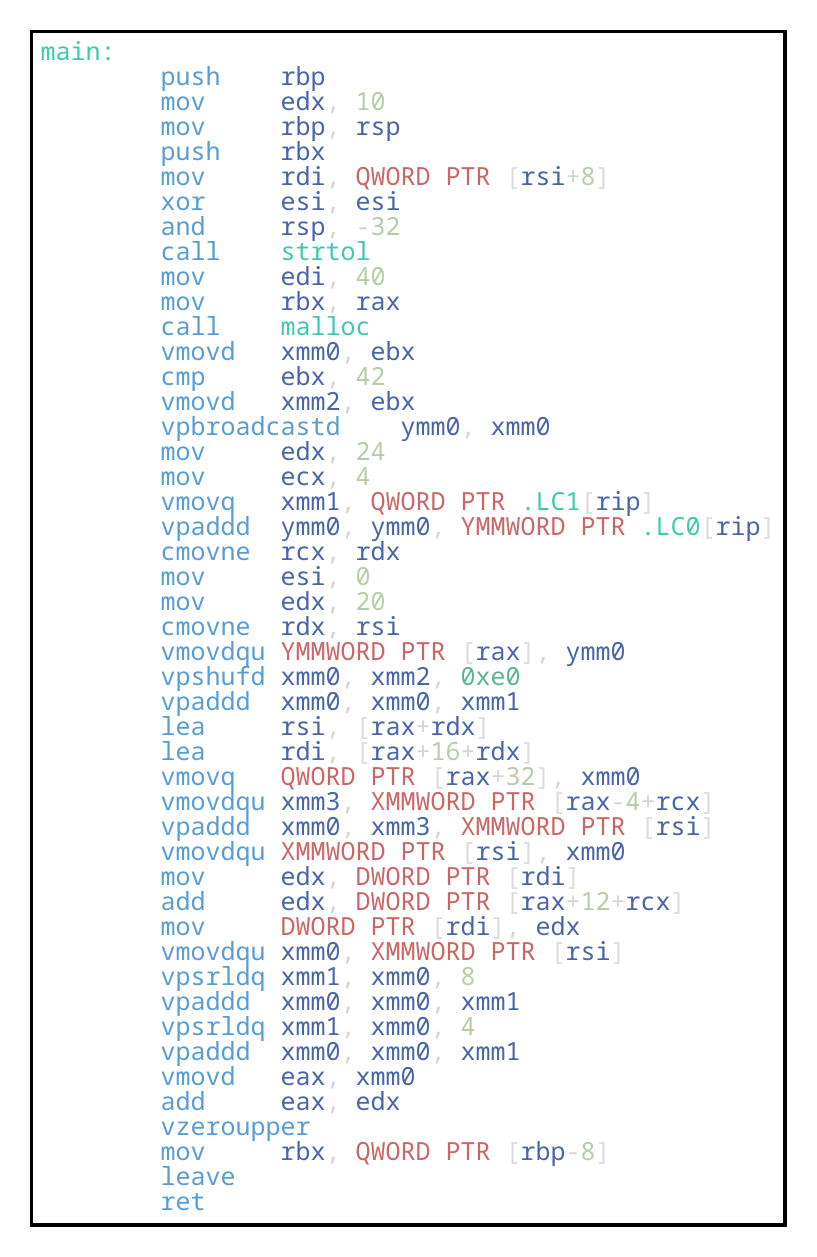}};
    \end{tikzpicture}
    \end{minipage}
    \vspace{-1em}
    \caption{Left side: example code using pointer movements where the points-to analysis inside GCC is unable to ensure non-aliasing accesses. Right side: code after \textit{adjunct} transformation where GCC is able statically decide non-aliasing using points-to analysis and integer analysis. Note the difference in the CFG of the resulting assembly code.}
    \label{points-to example}
\end{figure}
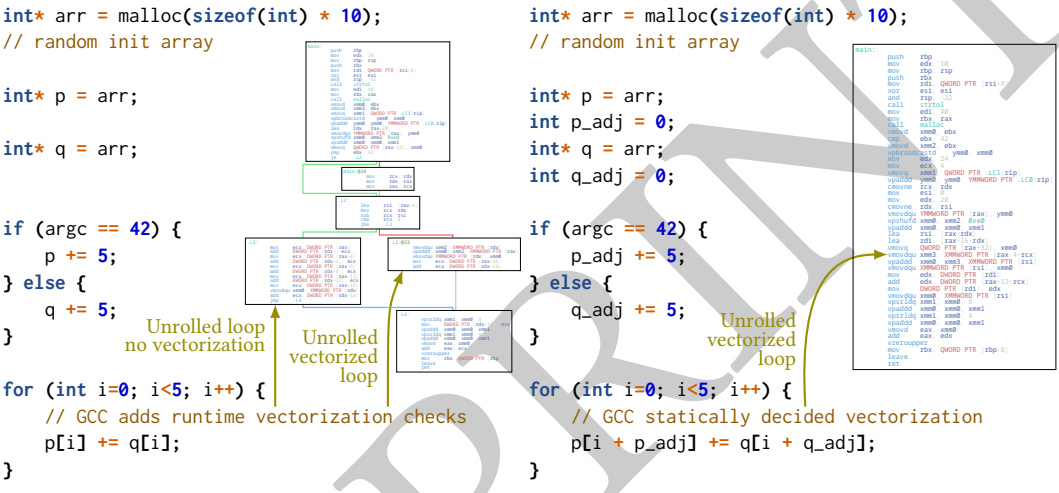

While points-to analysis represents a cutting-edge method widely integrated into modern compilers, its effectiveness in element-sensitive analysis, particularly concerning pointer movements, remains limited. This constraint poses challenges for loop optimization strategies like vectorization or parallelization, as points-to analysis struggles to grasp element-sensitive aliasing patterns. As illustrated in Figure~\ref{points-to example}, conventional points-to analysis, as implemented in GCC 13, fails to statically discern element-sensitive aliasing in code involving pointer movements. However, following our transformation, GCC successfully identifies the absence of access collisions. It is worth noting that even after the \textit{adjunct} transformation, GCC continues to leverage points-to analysis to recognize the independence between accessed elements. The shown example in Figure~\ref{points-to example} includes only a 5-iteration loop to more obviously present the missing vectorization check branch.

Figure~\ref{points-to example} illustrates that the transformation is also beneficial for dynamic allocations. In such cases, the base pointer must be stored somewhere to ensure proper memory deallocation. However, even with the base pointer known, the compiler often struggles to fully resolve all aliasing information.

Sui et al.~\cite{10.1145/3168364} proposed additional analysis techniques for pointer loops, enhancing field-sensitive and element-sensitive methods. However, their work does not address pointer movements, which remain a challenge for compilers, even advanced ones. Our transformation effectively eliminates pointer movements, allowing existing points-to analysis techniques, including advanced element-sensitive approaches, to be applied with greater efficacy. By incorporating points-to analysis with our approach, we enhance the depth of insights and enable advanced code optimization. Our methodology involves modifying the source code, rendering it compatible with various analyzers and optimizers. While our primary focus lies in data-centric frameworks, which often struggle with handling pointers, it is essential to note that our approach is not limited to this domain. The modified source code generated through our method seamlessly integrates with any existing or future frameworks, offering versatility and long-term applicability.

Scalar Evolution (SCEV)~\cite{absar2018scalar} and other loop analysis methods aim to scrutinize loops for potential optimizations and performance enhancements, including vectorization. However, as demonstrated in numerous examples, current implementations often struggle to identify such opportunities, particularly concerning pointers. Although there have been proposals for improving analysis techniques~\cite{kim2019code, belyaev2014improving, gunther2006profile, 10.1145/3168364}, finding techniques that are both safe and general enough remains challenging. Our transformation addresses this challenge by providing additional information to the compiler, enabling it to leverage existing SCEV techniques for performance improvements. With better analysis techniques, more patterns can be matched. Importantly, our transformation does not hinder the adoption of new techniques, as the array access pattern is well-established and commonly utilized, making it compatible with existing analyzers. While advanced analyzers may eventually incorporate pointer movements into their analyses, currently, this remains a significant challenge, and most analyzers do not account for it. Hence, we contend that pointer movement continues to pose a problem, and our transformation provides a valuable solution in the current landscape.

\subsection{Practical results} \label{practical_results}
\paragraph{PBKDF2}
To showcase the transformation in practice we test it on the code snippet from PBKDF2 first presented in Section~\ref{sec:intro}. We expect the inner loop $\circledColorSmall{2}{black}$ to be vectorized by every compiler. What we want to test is whether compilers are able to discover the independence between the iterations of the outer loop $\circledColorSmall{1}{black}$ and the effects that the \textit{adjunct} transformation has on it.

\noindent\begin{minipage}{0.515\linewidth}
\begin{minted}{C}
// snippet of the PBKDF2 implementation in OpenSSL
// int cplen, n
for (int i=0; i<cplen*n; i+=cplen) {
  for (int k=0; k<cplen; k++) {
    // no loop carried dependence
    p[k] = p[k] ^ data[k];
  }
  // each outer loop iteration works on different data
  p += cplen; 
}
\end{minted}
\end{minipage}
\hfill
\begin{minipage}{0.475\linewidth}
\begin{minted}{C}
// adjunct transformed version
// int cplen, n
for (int i=0; i<cplen*n; i+=cplen) {  @$\circledColorSmall{1}{black}$@
  for (int k=0; k<cplen; k++) {       @$\circledColorSmall{2}{black}$@
    // data adjunct removed for readability
    p[k + p_adj] = p[k + p_adj] ^ data[k];
  }
  // no pointer movement, the adjunct variable is used instead
  p_adj += cplen;
}
\end{minted}
\end{minipage}

\hfill

The code was run both with normal pointer movements (``no \textit{adjunct}'') and with the transformation (``with \textit{adjunct}'' variables). It was compiled with different compilers: GCC 12.2.2, Clang 15.0.6, and Polly (same Clang version). We also used the DaCe 0.14.1 data-centric framework to analyze and optimize the code with the resulting file being compiled with GCC. The code was instrumented with PAPI~\cite{1393} to measure the execution time and the number of instructions. For each test, 10 runs were executed and the median is reported, with the 95\% confidence interval being reported. For the number of instructions, the error range is negligible ($\pm 1000$ instructions) and not reported. Runs were done with \texttt{n = 100} and \texttt{cplen = $100 \cdot 10^6$}. As DaCe and Polly produce OpenMP code we report different values for different thread counts. We compare results with Polly as it is another compiler that offers automatic parallelization. Note that after the \textit{adjunct} transformation the code only has array accesses which is a pattern supported by polyhedral compilers.

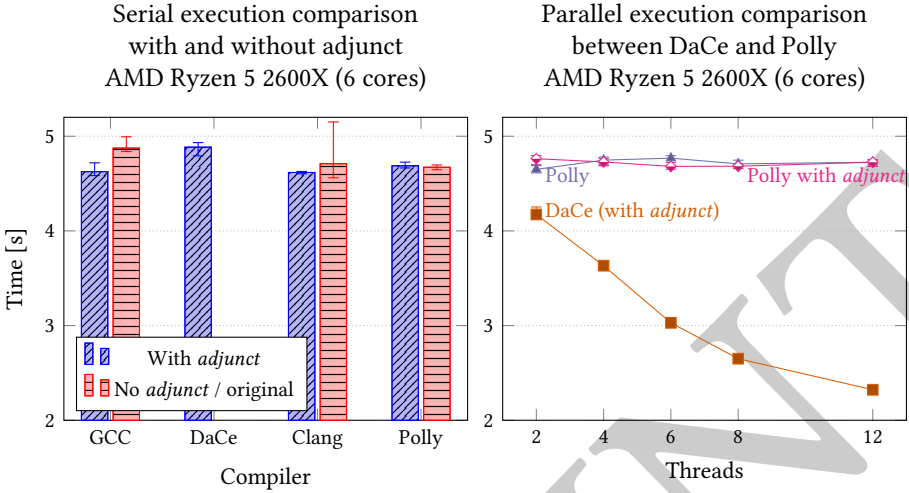
\begin{figure}[th]
    \centering
    \begin{tikzpicture}[baseline]
    \begin{axis}[
        xlabel={Compiler},
        ylabel={Time [s]},
        error bars/y dir=both,
        error bars/y explicit,
        title={Serial execution comparison\\with and without adjunct\\ AMD Ryzen 5 2600X (6 cores)},
        %x filter/.code={\pgfmathparse{#1/1000000}\pgfmathresult},
        %ytick scale label code/.code={},
        %y filter/.code={\pgfmathparse{#1/1000}\pgfmathresult},
        xtick={GCC, DaCe, Clang, Polly},
        symbolic x coords={GCC, DaCe, Clang, Polly},
        ymax=5.2,
        ymin=2,
        height=.4\textwidth,
        width=.5\textwidth,
        ybar,
        legend pos=south west,
        enlarge x limits=0.15,
        xtick align=inside,
    ]

        \addplot +[postaction={pattern=north east lines}] table [x index=0, y index=1, y error plus index=3, y error minus index=2] {simple_data/hi_twin.dat};
        
        \addplot +[postaction={pattern=horizontal lines}] table [x index=0, y index=1, y error plus index=3, y error minus index=2] {simple_data/hi_no_twin.dat};

        \legend{With \textit{adjunct}, No \textit{adjunct} / original}

    \end{axis}
    \end{tikzpicture}
    \begin{tikzpicture}[baseline]
    \begin{axis}[
        xlabel={Threads},
        ylabel={},
        error bars/y dir=both,
        error bars/y explicit,
        cycle list shift=1,
        title={Parallel execution comparison\\between DaCe and Polly\\ AMD Ryzen 5 2600X (6 cores)},
        %x filter/.code={\pgfmathparse{#1/1000000}\pgfmathresult},
        %ytick scale label code/.code={},
        %y filter/.code={\pgfmathparse{#1/1000}\pgfmathresult},
        xtick=data,
        ymax=5.2,
        ymin=2,
        height=.4\textwidth,
        width=.5\textwidth,
    ]

        \addplot table [x index=0, y index=1, y error plus index=3, y error minus index=2] {simple_data/simple_DaCe.dat}
        node[anchor=west,pos={0}, xshift=0, yshift=1] {\footnotesize DaCe (with \textit{adjunct})};

        \addplot table [x index=0, y index=1, y error plus index=3, y error minus index=2] {simple_data/simple_Polly.dat}
        node[anchor=west,pos={0}, xshift=0, yshift=-3] {\footnotesize Polly};

        \addplot table [x index=0, y index=1, y error plus index=3, y error minus index=2] {simple_data/simple_Polly_twin.dat}
        node[anchor=west,pos={3/5}, xshift=0, yshift=-4] {\footnotesize Polly with \textit{adjunct}};

    \end{axis}
    \end{tikzpicture}
    \vspace{-1em}
    \caption{Serial and parallel runtimes for a simplified code snippet extracted from the PBKDF2 implementation of OpenSSL. Note that DaCe can only analyze the code after using the \textit{adjunct} transformation.}
    \label{simple_results}
\end{figure}

Figure~\ref{simple_results} shows that the adjunct transformation gets a minimal improvement in single-threaded performance. While with multiple threads the data-centric compiler DaCe was able to identify the parallelization opportunity on loop $\circledColorSmall{1}{black}$. Polly however isn't able to identify the parallel loop either with or without the \textit{adjunct} transformation. 
This is the case because the array-style access of \texttt{p} is non-affine as it uses the \textit{p\_adj} variable which is non-affine w.r.t. loop iterators. Which is a requirement for the loop to be a SCoP (static control part). Polly only acts on SCoPs of programs. Note that as DaCe uses a data-centric IR which utilizes static and scope defined data containers (for arrays), pointer movements are not supported directly. The \textit{adjunct} transformation is a way to support pointer movements in the data-centric IR without the need to change its inner workings.

The number of instructions neither increase nor decrease with the \textit{adjunct}. As Polly is a collection of transformations on the LLVM-IR that is compiled using Clang the number of instructions is identical (Polly wasn't able to find any SCoPs hence no transformations were applied). Surprisingly DaCe has the same number of instructions as GCC. Even if it uses GCC to compile to machine instructions this is unexpected as DaCe applies multiple transformations to the code. This can be explained as even after the transformations the computation is mostly the same only with additional annotations for parallel execution.

\begin{table}[h]
\begin{tabular}{l l l}
    \textbf{Instructions [$10^6$]}  & No \textit{adjunct}  & With \textit{adjunct} \\
    \hline
    GCC                             & $4\,380$          & $4\,380$ \\
    Clang                           & $2\,974$          & $2\,974$ \\
    Polly                           & $2\,974$          & $2\,974$ \\
    DaCe                            & not supported     & $4\,380$ \\
    \hline
\end{tabular}
\caption{Number of instructions generated by different compilers of the code snippet taken from PBKDF2.}
\label{instructions_table}
\end{table}

\vspace{-1em}
This proves the value of the transformation, but also that the transformation benefits compilers with data-centric IR the most: they can leverage the improved analyzability of pointer movements. Data centric-paradigms otherwise represent pointers as indirection that is challenging - or even impossible - to handle. All compilers see a minimal improvement but they are less able to capitalize on the improved analyzability.

\paragraph{LZO (compression algorithms)}
Another instance where pointer movements within loops are prevalent is in compression algorithms. As an illustration, we applied our transformation to the compression function of the Lempel–Ziv–Oberhumer (LZO) algorithm~\cite{lzo}. This algorithm is commonly utilized within the Linux Kernel for file-systems and memory, supporting live compression/decompression operations. Notably, it finds applications in BTRFS~\cite{btrfs}, SquashFS~\cite{squashfs}, initramfs~\cite{initramfs}, zram~\cite{zram}, and zswap~\cite{zswap}.

Following a similar approach to previous benchmark studies of the LZO algorithm~\cite{6374778}, we executed multiple iterations of the algorithm, ensuring each run involved significant computation without increasing memory usage. The compression was performed on a 1 MB file with 2000 iterations and a block size of 256KB. As before, we report the median with a 95\% confidence interval using the same compiler versions. We compared the original implementation compiled with GCC against DaCe with the adjunct transformation. It is noteworthy that the DaCe code was executed with a single thread, as no significant parallelization opportunities were identified.

Our results indicate a slight improvement, thanks to the additional vectorization opportunities provided by the transformation. However, achieving more substantial improvements, along with potential parallelization opportunities, proves challenging due to loop carried dependencies which would require an algorithm redesign. Nevertheless, in compression algorithms, the adjunct transformation can facilitate modest yet discernible enhancements in runtime performance.

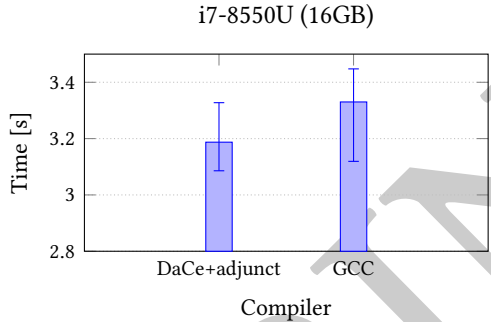
\begin{figure}[th]
    \centering
    \begin{tikzpicture}[baseline]
    \begin{axis}[
        xlabel={Compiler},
        ylabel={Time [s]},
        error bars/y dir=both,
        error bars/y explicit,
        title={i7-8550U (16GB)},
        %x filter/.code={\pgfmathparse{#1/1000000}\pgfmathresult},
        %ytick scale label code/.code={},
        %y filter/.code={\pgfmathparse{#1/1000}\pgfmathresult},
        xtick={DaCe+adjunct, GCC},
        symbolic x coords={DaCe+adjunct, GCC},
        ymax=3.5,
        ymin=2.8,
        enlarge x limits=1,
        height=.3\textwidth,
        width=.5\textwidth,
        ybar,
        xtick align=inside,
    ]

        \addplot table [x index=0, y index=1, y error plus index=3, y error minus index=2] {lzo/lzo.dat};

    \end{axis}
    \end{tikzpicture}

    \vspace{-1em}
    
    \caption{Serial runtime (single thread) of LZO compression algorithm. Comparison of original implementation compiled with GCC 12.2.2 and DaCe transformed version (including \textit{adjunct} transformation)}
    \label{lzo_results}
\end{figure}

\section{Limitations} \label{limitations}
The method we propose can provide compilers with better insights into pointer behavior, but only in certain scenarios. In this section, we go into detail as to the limitations of our approach. The main limitation is about the static decidability of which data container a pointer is pointing to. But we argue that in those cases an algorithm redesign would be needed for any performance improvement.

The necessity of static decidability arises from the utilization of DaCe as the compiler/framework for optimizations. Although DaCe achieves significant performance improvements, particularly in automatic parallelization, it relies on the ability to statically determine data containers. However, the \textit{adjunct} transformation itself encounters limitations primarily associated with higher-order pointers, as elaborated in Subsection~\ref{limit_double_pointers}. 

Another important aspect to highlight is the utility of our \textit{adjunct} transformation. Simply exposing additional information to the compiler does not guarantee a runtime performance improvement. In earlier examples, we demonstrated how the compiler generates fewer instructions or eliminates branches entirely. Additionally, using DaCe it created further parallelization opportunities. We discovered that these improvements are predominantly seen in codes where pointers are used as iterators. While this pattern is common in many codebases as we will discuss in Table~\ref{tab:codebases}, no immediate improvements will be offered to codes that do not exhibit this pattern.

\subsection{Static decidability}
A first assumption is that a pointer has to always point to a statically decidable data container in any given section of the code. This is required to be able to deterministically identify data accesses. Conditional reassignment introduces uncertainty as to which data container a pointer is pointing to. In such cases, our analysis backs off and does not generate an adjunct. However, unconditional or static pointer re-assignments are supported as they aren't introducing uncertainty as to which data container they point to. We can see two code snippets that show this difference in Figure~\ref{static_decidability}.

With this assumption, we achieve flawless aliasing detection, as the compiler possesses static knowledge of the data container associated with each pointer. Detecting accesses to the same data simply involves verifying if the data container and indices are identical. Leveraging the \textit{adjunct} transformation streamlines index comparison to a straightforward integer check, often allowing for static analysis if the abstract model is sufficiently precise.

It would be possible to expand the transformation to also support conditional pointers reassignment by creating adjunct versions for each container and replicating the branch decision for each subsequent access. However, this would create unsustainable code and memory replication requirements.

\begin{figure}[th]
\begin{minipage}{0.48\linewidth}
Decidable access - supported
\begin{minted}{C}
int* p;
p = a;
// now p always points to a
p[i] = 5; // accessing a with an offset of i

p = b;
// now p always points to b
p[i] = 2; // accessing b with an offset of i
\end{minted}
\end{minipage}
\hfill
\begin{minipage}{0.48\linewidth}
Undecidable access - not supported
\begin{minted}{C}
int* p;
if (...) { // unknown branch
  p = a;
} else {
  p = b;
}
// p could point to either a or b
p[i] = 5;  // undecidable data container
\end{minted}
\end{minipage}
\caption{Undecidable vs decidable pointer accesses}
\label{static_decidability}
\end{figure}

\subsection{Pointing to pointers}\label{limit_double_pointers}

\begin{figure}[h]
\begin{minipage}[t]{0.48\textwidth}
\begin{minted}{C}
int* a = new int[10];
                     @\tikzmark{ptr_init}@
int* b = new int[10];


int** p = new int*[2];
                     @\tikzmark{p_init}@


p[0] = a;
        @\tikzmark{adjunct_copy}@
p[1] = b;


if (...) {
    p++;
}

int x = (*p)[i] // -> either a[i] or b[i]
@\phantom@
\end{minted}
\begin{tikzpicture}[remember picture, overlay, olive]
    \draw[-latex] ([shift={(0,0)}] pic cs:ptr_init) -- node[midway, above] {\footnotesize adjunct for new pointers} +(4, 0);
    \draw[-latex] ([shift={(0,0)}] pic cs:p_init) -- node[midway, below] {\footnotesize Array of adjunct for new pointer array} +(4, 0);
    \draw[-latex] ([shift={(0,0)}] pic cs:adjunct_copy) -- node[midway, below, align=center] {\footnotesize Copy current adjunct position for \\ \footnotesize future accesses through \texttt{p}} +(5.81, 0);
\end{tikzpicture}
\end{minipage}
\hfill
\begin{minipage}[t]{0.48\textwidth}
\begin{minted}{C}
int* a = new int[10];
int a_adj = 0;
int* b = new int[10];
int b_adj = 0;

int** p = new int*[2];
int* p_adj_array = new int[2];
int p_adj = 0;

p[0] = a;
p_adj_array[0] = a_adj;
p[1] = b;
p_adj_array[1] = b_adj;

if (...) {
    p_adj++;
}
           @\tikzmark{access}@
int x = (p[p_adj])[i + p_adj_array[p_adj]];
@\phantom@                 @\tikzmark{array_access}@
\end{minted}
\begin{tikzpicture}[remember picture, overlay, olive]
    \draw[-latex] ([shift={(0,0)}] pic cs:access) -- node[pos=0, left] {\footnotesize Offset of \texttt{p} pointer} +(0.4, -0.15);
    
    \draw[-latex] ([shift={(1.5,0)}] pic cs:array_access) -- node[pos=0, right] {\footnotesize Offset of array element} +(-0.35, +0.25);
    
    \draw[-latex] ([shift={(0,0)}] pic cs:array_access) -- node[pos=0, left] {\footnotesize Requested index} +(0.35, +0.25);
    
    \draw[-latex] ([shift={(3.9,0.2)}] pic cs:access) -- node[pos=0, above, align=center] {\footnotesize Based on which element \\ \footnotesize we are accessing with \texttt{p}} +(0, -0.35);
\end{tikzpicture}
\end{minipage}
\vspace{-1em}
\caption{Example of adjunct transformation with arrays of pointers}
\label{adjunct_array}
\end{figure}
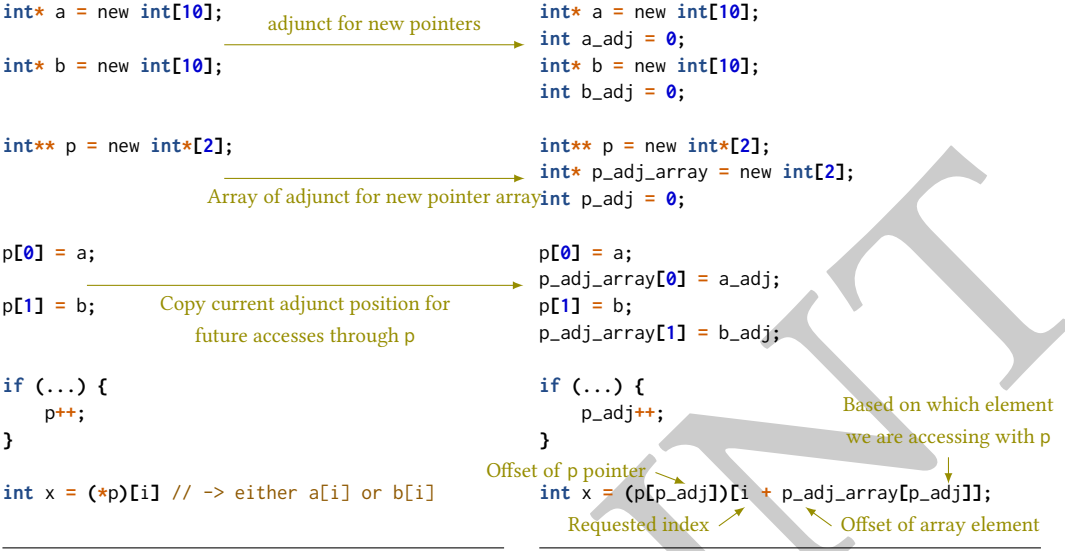

Double pointers - or higher order pointers - are supported by applying the transformation recursively, as can be seen in Figure~\ref{adjunct_double_pointers}. 
Note that the data container to which \texttt{p} is pointing to is statically known. As with simple pointers, we can trace the accesses $\circledColorSmall{2}{olive}$ and $\circledColorSmall{3}{olive}$ to container \texttt{a} ($\circledColorSmall{1}{olive}$) and get the correct \textit{adjunct} in this case. But the decidability issue remains, as we can see in Figure~\ref{adjunct_array} (left side) where the data container that is accessed is determined by the higher-order pointer location.

\begin{figure}[h]
\begin{minipage}{0.49\textwidth}
\begin{minted}{C}
int* a = new int[10];


int** p = &a; 


int x = (*p)[i] // -> a[i]
a++;
int y = (*p)[i] // -> a[i + 1]
\end{minted}
\end{minipage}
\hfill
\begin{minipage}{0.49\textwidth}
\begin{minted}{C}
int* a = new int[10];
int a_adj = 0;

int** p = &a; @$\circledColorSmall{1}{olive}$@
int p_adj = 0;

int x = (p[p_adj])[i + a_adj] @$\circledColorSmall{2}{olive}$@
a_adj++;
int y = (p[p_adj])[i + a_adj] @$\circledColorSmall{3}{olive}$@
\end{minted}
\end{minipage}
\caption{Example of adjunct transformation with double pointers}
\label{adjunct_double_pointers}
\end{figure}

The general case could be handled with arrays of \textit{adjunct}, as shown in Figure~\ref{adjunct_array} (right side). However, trying to leverage this approach on general codes would substantially increase the complexity of the code and consume a significant amount of memory. Note that this approach wasn't implemented because of the unsustainability of the transformation. Therefore, if additional assumptions can be made about the shape of the individual data containers pointed at, more targeted, efficient approaches can be used as explained in Section~\ref{specific_transformations}.

Although a multidimensional array shares the same memory representation as higher-order pointers, the analysis differs. Our aim is to eliminate pointer movements and replace them with more predictable array accesses. With multidimensional arrays, this pattern is inherent, and our transformation results in code similar to what is typically observed, as pointers are generally not relocated from multidimensional arrays. Prior proposals have suggested transformations to flatten multidimensional arrays into a single dimension~\cite{10.1145/2751205.2751248}. This approach was also included in GCC between version 4.3 and 4.8. The method is orthogonal to ours and can be applied before or after our transformation as it will not introduce additional pointer movements.

\subsection{Evaluating applicability}
In our investigation, we have identified the primary constraint of the \textit{adjunct} transformation: higher-order pointers. The requirement for statically decidable data containers is only a prerequisite of the DaCe framework and not of the \textit{adjunct} transformation itself. To gauge the broader viability of the transformation, we conducted an assessment of its utilization in extensive codebases, specifically focusing on the handling of higher-order pointers.

It is important to clarify our criteria for identifying non-applicable instances. We categorize higher-order pointer usage as constraining only when it potentially involves pointer iteration. Notably, we exempt instances where double pointers are employed to delegate memory allocation responsibilities to another function (refer to Figure~\ref{pointer_handoff}). Moreover, we exclude considerations of the \texttt{argv} parameter in main functions due to its typically diminutive size and minimal involvement in computationally intensive tasks. Our analysis considers application source code, omitting unit tests and examples. Furthermore, our scope is confined to C and C++ files. In the same way, we count single pointers as possible candidates for the \textit{adjunct} transformation.

\begin{figure}[h]
    \begin{minted}{C}
void init_pointer(char** p, int n) {
    *p = malloc(n);
    memset(*p, 0, n);
}

int main(int argc, char** argv) {
    char* buf;
    init_pointer(&buf, atoi(argv[1]));
    
    return 0;
}
    \end{minted}
    \caption{Example usage of double pointers to delegate memory allocation responsibility. This usage is fully supported by the \textit{adjunct} transformation.}
    \label{pointer_handoff}
\end{figure}

The outcomes of our investigation are presented in Table~\ref{tab:codebases}. Our examination encompassed OpenSSL~\cite{openssl}, notable for its cryptographic algorithm implementations, TurboBench~\cite{turbobench}, which hosts a plethora of compression algorithm implementations, and the Linux kernel~\cite{linux_kernel}.

\begin{table}[h]
\begin{tabular}{|r|r|r|r|}
     \hline
     \textbf{Code-base} & \textbf{LoC} & \textbf{\#applicable} & \textbf{\#non-applicable} \\
     \hline
     OpenSSL & 704,308 & 26,925 & 1,362 \\
     \hline
     TurboBench & 854,548 & 32,680 & 756\\
     \hline
     Linux kernel & 24,940,479 & 1,656,065 & 4,972\\
     \hline
\end{tabular}
\caption{Number of lines of code (LoC), number of applicable instances of the \textit{adjunct} transformation (\#applicable) and number of non applicable instances (\#non-applicable) in different code-bases (only C and C++).}
\label{tab:codebases}
\end{table}

Higher order pointers are indeed common in these codebases, as expected. However, their frequency is generally several orders of magnitude lower than the total code volume and notably even lower than single order pointer instances. It's essential to note that the provided statistics encompass pointers not exclusively used as iterators. This assessment offers a preliminary insight into the method's broader applicability, highlighting the prevalence of single pointers compared to their higher order counterparts in programs.

\section{From Generic to Specific - Refinements for Practical C Codes}
\label{specific_transformations}
We will show how our approach towards improving the analyzability of pointers used as iterators benefits real codes. However, we discovered that for such codes further refinements are needed such as additional transformations and support for pointers to external calls. All transformation in the following subsections were implemented and are done automatically by the compiler.

\subsection{Improving analyzability by restructuring data} \label{lil_transform_section}
As discussed in Section~\ref{limitations}, pointing to pointers is a more difficult to analyze scenario, as conditional pointer movements can lead to static undecidability regarding which data container is accessed. However, if additional information is known, the data structures themselves can be changed to make analysis easier. As this kind of transformation targets a general structure they do not only target a specific program but the entire set of programs that use that structure.

\iffalse
\begin{figure}[h]
    \begin{minted}{C}
int max_nnz = 27; // maximum non-zeros per row
int nrow = nx*ny*nz; // given as parameters
A->nnz_in_rows = new int[nrow]; // non-zeros per row
A->ptr_to_vals = new double*[nrow]; // start of values
A->ptr_to_inds = new int*[nrow]; // start of indicies
double * curvalptr = new double[max_nnz*nrow]; // values (contiguos list)
int * curindptr = new int[max_nnz*nrow]; // indicies (contiguos list)
for (...) { // iterate on every row
  // set where this row starts in the two contiguos arrays
  A->ptr_to_vals[currow] = curvalptr; @$\circledColorSmall{1}{teal}$@
  A->ptr_to_inds[currow] = curindptr; @$\circledColorSmall{2}{teal}$@
  int nnz_row = 0;
  for (...) { // iterate on every column
    *curvalptr = 27.0;
    *curindptr = curcol;
    curvalptr++;
    curindptr++;
    nnz_row++;
  }
  A->nnz_in_rows[currow] = nnz_row; // non-zeros for this row
}
    \end{minted}
    \caption{Initialization of a sparse matrix in LIL format taken from the HPCCG benchmark}
    \label{hpccg_init}
\end{figure}
\fi

\begin{figure}[h]
    \begin{minted}{C}
for (int i=0; i< nrow; i++) {
  double sum = 0.0; // accumulate sum for each row
  int cur_nnz = A->nnz_in_row[i]; // number of non-zeros per row
  for (int j=0; j<cur_nnz; j++) {
    sum += A->ptr_to_vals[i][j] @$\circledColorSmall{3}{teal}$@ // matrix element
             * x[A->ptr_to_inds[i][j]]; // vector element
  }
  y[i] = sum;
}
    \end{minted}
    \caption{Matrix-vector multiplication with a sparse matrix in LIL format taken from the HPCCG benchmark}
    \label{hpccg_access}
\end{figure}

An example of a structure transformation that we implemented is inside the Mantevo HPCCG benchmark. It uses sparse matrix storage, specifically the List of Lists (LIL) format. The matrix is stored using two lists for each row containing the column indices and the non-zero values. The row lists are stored as continuous data, and pointers to them are used to access the correct location based on the current row. In Figure~\ref{lil_transform}, we show the initialization code for a data structure from the HPCCG benchmark (modified for simplicity).

The matrix is accessed using those two lists, as it can be seen in Figure~\ref{hpccg_access} for a matrix-vector multiplication from the same benchmark. With the knowledge that there exists a statically known maximum number of non-zeroes per row, we can transform the list of lists into a matrix, as when the data is accessed (like in $\circledColorSmall{3}{teal}$) we see that the access resembles a 2D array. We transform the initialization of the sparse matrix to utilize a 2D-like structure instead of a contiguous list. The transformation is shown in Figure~\ref{lil_transform}. Thanks to this change we now have data stored in a more regular pattern (2D array). This enables compilers to apply standard transformations and optimizations on 2D containers.

\begin{figure*}[h]
    \begin{minipage}{0.49\textwidth}
        \begin{minted}{C}
double * curvalptr = new double[max_nnz*nrow];
int * curindptr = new int[max_nnz*nrow];
for (...) { // for every row
  A->ptr_to_vals[currow] = curvalptr; @$\circledColorSmall{1}{teal}$@

  A->ptr_to_inds[currow] = curindptr; @$\circledColorSmall{2}{teal}$@

  for (...) { // for every value
    *curvalptr = 27.0;
    *curindptr = curcol;
    
    curvalptr++;
    curindptr++;
  }
}
        \end{minted}
    \end{minipage}
    \hfill
    \begin{minipage}{0.49\textwidth}
        \begin{minted}{C}
// do not allocate curvalptr or curindptr

for (...) { // for every row
  A->ptr_to_vals[currow] = new double[max_nnz];
  int vals_adj = 0;
  A->ptr_to_inds[currow] = new int[max_nnz];
  int inds_adj = 0;
  for (...) { // for every value
    A->ptr_to_vals[currow][vals_adj] = 27.0;
    A->ptr_to_inds[currow][inds_adj] = curcol;
    
    vals_adj++;
    inds_adj++;
  }
}
        \end{minted}
    \end{minipage}

    \caption{Initialization of a sparse matrix in LIL format extracted from the Mantevo HPCCG benchmark. The original code is on the left, while the transformed code is on the right.}
    \label{lil_transform}
\end{figure*}

The transformation is done first by finding two pointer assignments working on the same struct inside the same scope ($\circledColorSmall{1}{teal}$ and $\circledColorSmall{2}{teal}$). We use a map to relate the right side of the assignment with the left (\texttt{curvalptr} to \texttt{A->ptr\_to\_vals[currow]}). We then create an \textit{adjunct} and transform the code in a similar way to the general pointer transformation. Using the previously created map we find all pointer accesses that must be transformed and substitute them with array access using the \textit{adjunct}. In the same way, we find pointer arithmetic operations and substitute the pointer with the \textit{adjunct}.

The benefit of using such specific pattern-based transformations is that they can be used in conjunction with our general algorithm, and therefore be integrated into the same workflow. This creates the possibility to efficiently handle a wide range of other access patterns without the need for separate workflows or manual code modification.

It's important to note that for multidimensional arrays, allocating a continuous buffer and then splitting it by storing pointers introduces uncertainty for static analysis, as explained in Section~\ref{limit_double_pointers}. Such patterns are unsupported by our current method. However, our LIL transformation automatically converts this pattern into a 2D array. In other scenarios, implementing additional transformations may be necessary to align with the limitations of our method.

\subsection{Handling external calls with pointer arguments}\label{ext_calls}
Some parts of a program, such as external library calls are outside the scope of static compiler analysis. These must therefore be considered black boxes. A common pattern in C when calling external functions is to use pointers as arguments for inputs and outputs. One such example is the \texttt{memcpy} function from the C standard library. The function with the following signature:
\begin{minted}{C}
void* memcpy(void* destination, const void* source, size_t num);
\end{minted}
takes two pointers and an integer \texttt{num} and copies \texttt{num} bytes from the \texttt{source} pointer to the \texttt{destination}.

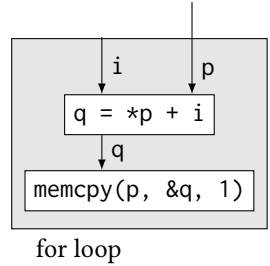
\begin{figure}[h]
\begin{minipage}{0.48\textwidth}
\begin{minted}{C}
char* p = new char[1];
*p = 0;
for (int i=0; i<n; i++) {
  char q = *p + i;   // use *p from previous iteration
  memcpy(p, &q, 1); // equal to *p = q
}
\end{minted}

\centering
\begin{tikzpicture}
        \node[anchor=north west] at (-1.7, 3) {With \texttt{memcpy} dependency};
        \draw[draw=black, fill=boxbg] (-1.7, 2) rectangle ++(3.4, -2.5);
        
        \node[rectangle, draw, fill=white] (assignment) at (0, 1) {\texttt{q = *p + i}};
        \node[rectangle, draw, fill=white] (memcpy) at (0, 0) {\texttt{memcpy(p, \&q, 1)}};
    
        \draw[-latex] ([shift={(-0.5,0)}] assignment.south) -- node[midway, right] {\texttt{q}} ([shift={(-0.5,0)}] memcpy.north);
        \draw[-latex] ([shift={(0.7,1.75)}] assignment.south) -- node[pos={3/4}, right] {\texttt{p}} ([shift={(0.7,0)}] assignment.north);
    
        \draw[-latex,] ([shift={(-0.5,0.75)}] assignment.north) -- node[midway, right] {\texttt{i}} ([shift={(-0.5,0)}] assignment.north);

        \draw[-latex, dashed] ([shift={(0.7,0)}] memcpy.south) -- +(0, -0.5) -- +(1.2, -0.5) -- +(1.2, 2.5) -- +(0.05, 2.5);
    
        \node[anchor=north west] at (-1.5, -0.5) {for loop};
    \end{tikzpicture}
\end{minipage}
\hfill
\begin{minipage}{0.48\textwidth}
\begin{minted}{C}
int* q = new char[1];
*q = 42;
for (int i=0; i<n; i++) {
  char p = 0;
  memcpy(&p, q, 1); // equal to p = *q
  data[i] = p + i;  // data[i] = 42 + i;
}
\end{minted}

\centering
\begin{tikzpicture}
        \node[anchor=north west] at (-1.7, 3) {Without \texttt{memcpy} dependency};
        \draw[draw=black, fill=boxbg] (-1.7, 2) rectangle ++(3.4, -2.5);
        
        \node[rectangle, draw, fill=white] (assignment) at (0, 1) {\texttt{q = *p + i}};
        \node[rectangle, draw, fill=white] (memcpy) at (0, 0) {\texttt{memcpy(p, \&q, 1)}};
    
        \draw[-latex] ([shift={(-0.5,0)}] assignment.south) -- node[midway, right] {\texttt{q}} ([shift={(-0.5,0)}] memcpy.north);
        \draw[-latex] ([shift={(0.7,1.75)}] assignment.south) -- node[pos={3/4}, right] {\texttt{p}} ([shift={(0.7,0)}] assignment.north);
    
        \draw[-latex,] ([shift={(-0.5,0.75)}] assignment.north) -- node[midway, right] {\texttt{i}} ([shift={(-0.5,0)}] assignment.north);
    
        \node[anchor=north west] at (-1.5, -0.5) {for loop};
    \end{tikzpicture}
\end{minipage}

\caption{Left side: \texttt{memcpy} creates a dependency between loop iterations. Right side: \texttt{p} is a loop local variable and \texttt{memcpy} is used to assign values to it.}
\label{memcpy_example}
\end{figure}

In the previous section, we discussed the possible modifications and movement of pointers and their impact on decidability. In the example of \texttt{memcpy} pointers are not modified. In C arguments are passed by copy and not by reference therefore callee modifications of arguments are not propagated to the caller. The same argument applies to pointers. What is modified in those cases is the content of the pointer, hence the elements of the data container that the pointer is pointing to. A function taking a pointer as an argument can modify its contents. This poses a challenge as data dependencies can no longer be tracked. In Figure~\ref{memcpy_example}, we use \texttt{memcpy} instead of a simple assignment.

On the left side of the figure the data in \texttt{p} depends on the value previously written to it by \texttt{memcpy} (dependency shown by the dashed arrow). This enforces that the loop must be executed in order. If we remove this dependency (dashed arrow) then loop iterations would be independent. This would permit the compiler to vectorize the loop producing the incorrect result.

We previously assumed that the second argument of \texttt{memcpy} is only read, but this is an insight gained by looking at the implementation of \texttt{memcpy} and not black box analysis. Without outside information, no compiler framework can overcome this issue. In the right side of Figure~\ref{memcpy_example} \texttt{q} is only read from but the compiler cannot infer this. This creates a spurious data dependency between iterations and prevents the loop from being parallelized.

While it is good practice to make pointer arguments that are read-only \texttt{const}, not every function or library does so. Our solution is to use whitelists for often-used library calls - manually annotating library functions to specify if a pointer is write-only, read-write, or read-only. This whitelist could become a database of known functions that could even be automatically populated through compiler analysis, as it only needs to be performed once for any given library implementation. Annotations only provide information about which data container was read or written. We, therefore, assume conservatively that the entire data container could be read or written.

\subsection{Stateful external calls}\label{stateful_calls}
Some libraries store an internal state that is opaque to the program calling functions provided by the library --- great examples being MPI or OpenGL. One such library is a part of OpenSSL, namely HMAC (Hash Message Authentication Code) --- it uses an internal state to store hashes and keys between calls. This data is saved in an internal data structure and a pointer to it is provided to calling functions. This pointer (called context in this implementation) is never modified by the user code, but only by library calls.

\begin{minted}{C}
HMAC_CTX *hctx = HMAC_CTX_new();  @$\circledColorSmall{1}{orange}$@
for (...) {
  HMAC_CTX_copy(hctx, base_hctx); @$\circledColorSmall{2}{orange}$@
  // do operations with hctx
}
HMAC_CTX_free(hctx);
\end{minted}

In the example above, without any additional information, it appears that the \texttt{hctx} pointer is initialized outside the loop ($\circledColorSmall{1}{orange}$), and that the call to \texttt{HMAC\_CTX\_copy} within the loop ($\circledColorSmall{2}{orange}$) could read and write to \texttt{hctx}. This creates a data dependency between loop iterations. In reality,  \texttt{HMAC\_CTX\_copy} does not read the data of the first argument (\texttt{hctx}) --- that value is only written to and effectively used as a local variable within the scope of the loop. Therefore, by marking the dependency on (\texttt{hctx}) in \texttt{HMAC\_CTX\_copy} as write-only and noticing that it is never used afterwards, the loop can be parallelized. This approach essentially performs an escape analysis~\cite{10.1145/1064979.1064996}, identifying that the variable within the loop does not escape the thread executing each iteration.

\section{Evaluation} \label{results}

To evaluate our transformation we extended the DaCe~\cite{dace,c2dace} pipeline by adding the pointer disaggregation transformation explained in Section~\ref{implementation} and the specific transformations explained in Section~\ref{specific_transformations}. The approach described in Section~\ref{ext_calls} and Section~\ref{stateful_calls} is implemented by adding a whitelist of additional dependencies in the data-centric IR (intermediate representation) used by DaCe.

As detailed and demonstrated in Section~\ref{practical_results}, while incremental performance enhancements are attainable across various compilers, the constrained scope of vectorization inhibits substantial performance gains. Consequently, we prioritize automatic parallelization compilers like Polly and DaCe, which hold promise in uncovering novel opportunities facilitated by the transformations implemented. Specifically, we direct our attention to DaCe due to its inherent data-centric approach, which offers superior analysis of data dependencies.

We measured performance on a dual-socket 2x6 core (2x12 threads) Intel Xeon X5670 @ 2.93GHz with 48GB of RAM. We compare the  DaCe data-centric framework (using GCC 12.1.1 as a backend compiler) against GCC and Polly (using Clang 15.0.6). We report the median of 10 runs with a confidence interval of 95\%. OpenMP is used for parallel execution. Integrating the \textit{adjunct} transformation into Polly did not reveal any additional Static Control Parts (SCoPs), and consequently, no further parallelization opportunities were identified. Hence, we provide the runtime results of Polly without the \textit{adjunct} transformation. We specifically chose to compare with Polly due to its widespread availability as part of the LLVM/Clang suite, offering some level of automatic parallelization capabilities. Our evaluation was conducted on the Mantevo HPCCG benchmark and the OpenSSL PBKDF2 implementation, with the results presented below.

\begin{figure}[th]
	\centering
	\begin{tikzpicture}[-latex]
        \draw[draw=black, fill=olive!25] (-4, 7.5) rectangle ++(8, -3);
        \draw[draw=black, fill=olive!50] (-4.5, 7.5) rectangle ++(0.5, -3);
        
        \draw[draw=black, fill=green!25] (-4, 4.5) rectangle ++(8, -3);
        \draw[draw=black, fill=green!50] (-4.5, 4.5) rectangle ++(0.5, -3);
        
        \draw[draw=black, fill=black!10] (-4,1.5) rectangle ++(8,-4.1);
        \draw[draw=black, fill=black!30] (-4.5,1.5) rectangle ++(0.5,-4.1);

        \node[anchor=north west, rotate=90] at (-4.5, 4.5) {\textbf{\textit{C transformation}}};
        \node[anchor=north west, rotate=90] at (-4.5, 1.5) {
            \textbf{\textit{Data centric}}
        };
        \node[anchor=north west, rotate=90] at (-4.5, -2.6) {
            \textbf{\textit{Hardware specific}}
        };
 
		\node[draw, fill=white] at (0,7) (input) {Input C file};
		\node[draw, fill=white] at (0,4.5) (ast) {C AST};
		\node[draw, fill=white] at (0,3) (sdfg) {Data flow graph};
  
		\node[draw, fill=white, align=center] at (0,0) (output) {Parallel C++ code\\with OpenMP};

        \node[draw, fill=white, align=center, rounded corners=5pt] at (0, 1.5) (hwgen) {
            CPU\\
            code generator
        };

        \node[draw=black!50, text=black!50, fill=white, align=center, rounded corners=5pt] at (-2.5, 1.5) (hwfpga) {FPGA\\code generator};
        \node[draw=black!50, text=black!50, fill=white, align=center, rounded corners=5pt] at (2.5, 1.5) (hwcuda) {CUDA\\code generator};
        
        \node[draw, fill=white, align=center] at (0,-2) (asm) {Machine\\instructions};

		\path (input.south)	edge [thick] node [midway, right] {C parsing} (ast.north) ;

		\path ($(ast.west)+(0, 0.2em)$) edge [thick, loop left, looseness=10, out=180, in=90]
            node [pos=0.8, above=-1em, align=left, anchor=south east] {
                AST transformations\\
                including \textit{adjunct}
        } ($(ast.north west)+(0.6em, 0)$);
  
		\path (ast.south) edge [thick, align=left]
            node [midway, right] {
                Transformation to\\
                data centric IR
        } (sdfg.north);
            
        \path (sdfg.south west) edge [thick, loop left, looseness=5]
            node [pos=0.75, above=0.2em, xshift=-0.2em, align=left] {
                Data-centric\\
                transformations
        } (sdfg.north west);
        
		\path (sdfg.south) edge [thick] (hwgen.north);
        \path (hwgen.south) edge [thick] (output.north);

        \path ($(sdfg.south)+(-1em, 0)$) edge [draw=black!50, thick, out=-90, in=90] (hwfpga.north);
        \path ($(sdfg.south)+(1em, 0)$) edge [draw=black!50, thick, out=-90, in=90] (hwcuda.north);
            
        \path (output.south) edge [thick] node [midway, right, align=left] {C++\\compilation} (asm.north);
        
	\end{tikzpicture}
	\caption{Workflow for C program optimization including \textit{adjunct} transformation and leveraging the DaCe data-centric framework.}
\end{figure}
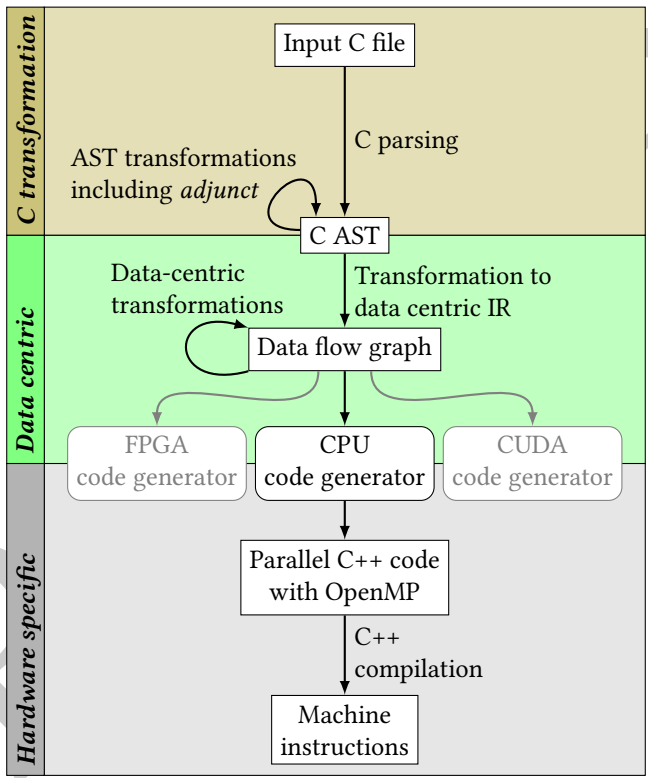

\subsection{OpenSSL PBKDF2}
The Password Based Key Derivation Function 2 (PBKDF2) is a derivation function that derives a cryptographically secure key from a password. PBKDF2 applies a Hash-based Message Authentication Code (HMAC) function multiple times to a provided password and salt. Based on the required length of the key this process is repeated multiple times. A summary of the process is provided in Figure~\ref{pbkdf_scheme}. We can notice that each single $B_i$ blocks can be computed independently of any other.

\newcommand\mdoubleplus{\mathbin{+\mkern-5mu+}}
\begin{figure}[th]
\begin{align*}
   key   &= PBKDF2(HMAC, password, salt, iterations, keyLen) \\
         &= B_0 \mdoubleplus B_1 \mdoubleplus \ldots \mdoubleplus B_{keyLen/hlen} \\[0.8em]
   B_i   &= U_0^i \oplus U_1^i \oplus \ldots \oplus U_{iterations}^i \\[0.8em]
   U_0^i &= HMAC(password, salt + i) \\
   U_1^i &= HMAC(password, U_0^i) \\
         &\vdots\\
   U_{iterations}^i &= HMAC(Password, U_{iterations - 1}^i)
\end{align*}
\caption{\texttt{PBKDF2} scheme. In this representation, $\mdoubleplus$ is the string concatenation, $\oplus$ is the XOR operator and $hlen$ is the output length of the HMAC function.}
\label{pbkdf_scheme}
\end{figure}

\begin{figure}[th]
\centering
\begin{tikzpicture}
\begin{groupplot}[
    group style={
        group size=2 by 2,  % 2x2 grid layout
        vertical sep=1cm,   % vertical separation between plots
        horizontal sep=1cm, % horizontal separation between plots
        ylabels at=edge left,% position the y-axis labels on the leftmost plots
        xlabels at=edge bottom,% position the x-axis labels on the bottom plots
    },
    every axis title/.style={at={(0,1)},anchor=south west},% Title position
    width=7.5cm,              % Width of each plot
    height=4.8cm,             % Height of each plot
    clip=false,
]

\iffalse
% Plot 1
\nextgroupplot[
    xlabel={},
    ylabel={Time [s]},
    error bars/y dir=both,
    error bars/y explicit,
    cycle list shift=1,
    title={\#blks=2},
    xtick=data,
    clip=false,
]
\addplot table [x index=0, y index=1, y error plus index=3, y error minus index=2] {pbkdf_data_2/pbkdf_dace_2.dat} node[anchor=south,pos={14/24}, xshift=0, yshift=0] {\footnotesize DaCe};

\addplot table [x index=0, y index=1, y error plus index=3, y error minus index=2] {pbkdf_data_2/pbkdf_polly_2.dat} node[anchor=east,pos={1}, xshift=0, yshift=-5] {\footnotesize Polly};

\addplot table [x index=0, y index=1, y error plus index=3, y error minus index=2] {pbkdf_data_2/pbkdf_parallel_2.dat} node[anchor=east,pos={1}, xshift=0, yshift=-4] {\footnotesize Manual OpenMP};

\addplot table [x index=0, y index=1, y error plus index=3, y error minus index=2] {pbkdf_data_2/pbkdf_fastpbkdf2_2.dat} node[anchor=west,pos={12/24}, xshift=0, yshift=-4] {\footnotesize FastPBKDF2};

\draw[very thick, pen colour={p_green}, decorate,decoration={calligraphic brace, mirror, amplitude=3pt}] (4.1, 1.3) -- node[text=black, right=0.4em, fill=white] {\footnotesize 1.9x} (4.1, 2.52);

\draw[very thick, pen colour={p_orange}, decorate,decoration={calligraphic brace, mirror, amplitude=3pt}] (4.1, 3.98) -- node[text=black, right=0.4em, fill=white] {\footnotesize 1.8x} (4.1, 7.29);

\fi

% Plot 2
\nextgroupplot[
    xlabel={},
    ylabel={Time [s]},
    error bars/y dir=both,
    error bars/y explicit,
    cycle list shift=1,
    title={\#blks=4},
    xtick=data,
    clip=false,
]

\addplot table [x index=0, y index=1, y error plus index=3, y error minus index=2] {pbkdf_data_2/pbkdf_dace_4.dat} node[anchor=west,pos={10/24}, xshift=4, yshift=7] {\footnotesize DaCe};

\addplot table [x index=0, y index=1, y error plus index=3, y error minus index=2] {pbkdf_data_2/pbkdf_polly_4.dat} node[anchor=east,pos={1}, xshift=0, yshift=-5] {\footnotesize Polly};

\addplot table [x index=0, y index=1, y error plus index=3, y error minus index=2] {pbkdf_data_2/pbkdf_parallel_4.dat} node[anchor=east,pos={1}, xshift=0, yshift=-5] {\footnotesize Manual OpenMP};

\addplot table [x index=0, y index=1, y error plus index=3, y error minus index=2] {pbkdf_data_2/pbkdf_fastpbkdf2_4.dat} node[anchor=west,pos={13/24}, xshift=0, yshift=-4] {\footnotesize FastPBKDF2};

\draw[very thick, pen colour={p_green}, decorate,decoration={calligraphic brace, amplitude=3pt}] (0.9, 1.3) -- node[text=black, left=0.4em, fill=white] {\footnotesize 3.7x} (0.9, 4.79);

\draw[very thick, pen colour={p_orange}, decorate,decoration={calligraphic brace, mirror, amplitude=3pt}] (4.06, 4.08) -- node[text=black, right=0.4em, fill=white] {\footnotesize 3.5x} (4.06, 14.25);

% Plot 3
\nextgroupplot[
    xlabel={},
    ylabel={},
    error bars/y dir=both,
    error bars/y explicit,
    cycle list shift=1,
    title={\#blks=8},
    xtick=data,
    clip=false,
]
\addplot table [x index=0, y index=1, y error plus index=3, y error minus index=2] {pbkdf_data_2/pbkdf_dace_8.dat} node[anchor=west,pos={6/24}, xshift=0, yshift=0] {\footnotesize DaCe};

\addplot table [x index=0, y index=1, y error plus index=3, y error minus index=2] {pbkdf_data_2/pbkdf_polly_8.dat}
node[anchor=east,pos={1}, xshift=0, yshift=-5] {\footnotesize Polly};

\addplot table [x index=0, y index=1, y error plus index=3, y error minus index=2] {pbkdf_data_2/pbkdf_parallel_8.dat}
node[anchor=west,pos={18/24}, xshift=0, yshift=5] {\footnotesize Manual OpenMP};

\addplot table [x index=0, y index=1, y error plus index=3, y error minus index=2] {pbkdf_data_2/pbkdf_fastpbkdf2_8.dat}
node[anchor=west,pos={14/24}, xshift=0, yshift=-4] {\footnotesize FastPBKDF2};

\draw[very thick, pen colour={p_orange}, decorate,decoration={calligraphic brace, mirror, amplitude=3pt}] (24.5, 4.5) -- node[text=black, right=0.4em, fill=white] {\footnotesize 6.2x} (24.5, 27.8);

\draw[very thick, pen colour={p_green}, decorate,decoration={calligraphic brace, mirror, amplitude=3pt}] (25, 1.3) -- node[text=black, right=0.4em, fill=white] {\footnotesize 7.2x} (25, 9.35);

% Plot 4
\nextgroupplot[
    xlabel={Threads},
    ylabel={Time [s]},
    error bars/y dir=both,
    error bars/y explicit,
    cycle list shift=1,
    title={\#blks=16},
    xtick=data,
    clip=false,
]
\addplot table [x index=0, y index=1, y error plus index=3, y error minus index=2] {pbkdf_data_2/pbkdf_dace_16.dat} node[anchor=west,pos={8/24}, xshift=0, yshift=0] {\footnotesize DaCe};

\addplot table [x index=0, y index=1, y error plus index=3, y error minus index=2] {pbkdf_data_2/pbkdf_polly_16.dat} node[anchor=east,pos={1}, xshift=0, yshift=-5] {\footnotesize Polly};

\addplot table [x index=0, y index=1, y error plus index=3, y error minus index=2] {pbkdf_data_2/pbkdf_parallel_16.dat} node[anchor=west,pos={16/24}, xshift=0, yshift=0] {\footnotesize Manual OpenMP};

\addplot table [x index=0, y index=1, y error plus index=3, y error minus index=2] {pbkdf_data_2/pbkdf_fastpbkdf2_16.dat} node[anchor=west,pos={16/24}, xshift=0, yshift=-4] {\footnotesize FastPBKDF2};

\draw[very thick, pen colour={p_orange}, decorate,decoration={calligraphic brace, mirror, amplitude=3pt}] (24.5, 6.26) -- node[text=black, right=0.4em, fill=white] {\footnotesize 8.9x} (24.5, 56);

\draw[very thick, pen colour={p_green}, decorate,decoration={calligraphic brace, mirror, amplitude=3pt}] (25, 2.1) -- node[text=black, right=0.4em, fill=white] {\footnotesize 8.9x} (25, 18.65);

% Plot 5
\nextgroupplot[
    xlabel={Threads},
    ylabel={},
    error bars/y dir=both,
    error bars/y explicit,
    cycle list shift=1,
    title={\#blks=24},
    xtick=data,
    clip=false,
]
\addplot table [x index=0, y index=1, y error plus index=3, y error minus index=2] {pbkdf_data/pbkdf_dace_24.dat}
        node[anchor=west,pos={8/24}, xshift=0, yshift=0] {\footnotesize DaCe};

        \addplot table [x index=0, y index=1, y error plus index=3, y error minus index=2] {pbkdf_data/pbkdf_polly_24.dat}
        node[anchor=east,pos={1}, xshift=0, yshift=-5] {\footnotesize Polly};

        \addplot table [x index=0, y index=1, y error plus index=3, y error minus index=2] {pbkdf_data/pbkdf_parallel_24.dat}
        node[anchor=west,pos={16/24}, xshift=0, yshift=0] {\footnotesize Manual OpenMP};
        
        \addplot table [x index=0, y index=1, y error plus index=3, y error minus index=2] {pbkdf_data/pbkdf_fastpbkdf2_24.dat}
        node[anchor=north,pos={12/24}, xshift=0, yshift=0] {\footnotesize FastPBKDF2};

        \draw[very thick, pen colour={p_green}, decorate,decoration={calligraphic brace, mirror, amplitude=5pt}] (24.7, 2.875) -- node[text=black, right=0.4em, fill=white] {\footnotesize 9.7x} (24.7, 27.9);

        \draw[very thick, pen colour={p_orange}, decorate,decoration={calligraphic brace, mirror, amplitude=5pt}] (24.3, 7.75) -- node[text=black, right=0.4em, fill=white] {\footnotesize 10.7x} (24.3, 83.3);

\end{groupplot}

\node (title) at ($(group c1r2.center)!0.5!(group c2r2.center)+(0,-2.8cm)$) {PBKDF2 with various block sizes};

\end{tikzpicture}

\caption{Runtime of various parallelized PBKDF2 implementations using different block counts (block count determines the problem size)  and using multiple thread counts. Increasing the thread count beyond the number of blocks would be ineffective as the work could not be meaningfully divided among threads.}
\label{openssl_graph_additional}
\end{figure}

The implementation inside OpenSSL does not include parallel options. We applied our pipeline to the OpenSSL code to auto-parallelize the code. We compare the runtime against Polly and a manually parallel version that was written using the OpenSSL implementation as a starting point. Only the PBKDF2 algorithm was analyzed, all HMAC calls are treated as external calls and use the standard OpenSSL implementation. For completeness, we compare it with FastPBKDF2~\cite{fastpbkdf}, an implementation of the same algorithm that has been developed for parallelism from the ground up.
We used SHA1 as the HMAC function, $5 \cdot 10^6$ iterations, and an output key size of 480 bytes (or 24 blocks), and summarize our results in  Figure~\ref{openssl_graph_additional}.

We then varied the problem size, represented by the number of blocks. The results (Figure~\ref{openssl_graph_additional}) show a consistent trend across all experiments. Our approach was able to obtain comparable results to the manually parallelized version, providing at most a 10.7x improvement with 24 threads. Polly wasn't able to identify any parallel opportunities due to the challenges of analyzing pointers. The FastPBKDF2 implementation is considerably faster than any OpenSSL equivalent, mainly due to the serial runtime of FastPBKDF2 version being 3 times faster, 27.7s versus 84.5s for OpenSSL. This reinforces the idea that tools can provide significant performance increases and find opportunities for parallelism, but nothing surpasses finding a better algorithm altogether.

\subsection{Mantevo HPCCG}
The Mantevo HPCCG benchmark computes the conjugate gradient on a sparse matrix. The sparse matrix is stored using the List of Lists (LIL) format. We leverage the specific transformation explained in Section~\ref{lil_transform_section} to improve analyzability. The LIL transformation is applied once only to the input file for DaCe. We compare our pipeline against Polly with the same preprocessed input. The baseline is obtained using the original benchmark with OpenMP enabled.

\begin{figure}[h]
    \begin{tikzpicture}
\begin{groupplot}[
    group style={
        group size=2 by 2,  % 2x2 grid layout
        vertical sep=1cm,   % vertical separation between plots
        horizontal sep=1cm, % horizontal separation between plots
        ylabels at=edge left,% position the y-axis labels on the leftmost plots
        xlabels at=edge bottom,% position the x-axis labels on the bottom plots
    },
    every axis title/.style={at={(0,1)},anchor=south west},% Title position
    width=7.5cm,              % Width of each plot
    height=4.8cm,             % Height of each plot
    clip=false,
]
% Plot 1
\nextgroupplot[
    xlabel={},
    ylabel={Time [s]},
    error bars/y dir=both,
    error bars/y explicit,
    cycle list shift=1,
    title={Size $100^3$},
    xtick=data,
    clip=false,
    ymin=0,
    xmin=-2,
    xmax=26,
]
\addplot table [x index=0, y index=1, y error plus index=3, y error minus index=2] {hpccg_data/hpccg_dace_100.dat} node[anchor=east,pos={1}, xshift=0, yshift=5] {\footnotesize DaCe};

\addplot table [x index=0, y index=1, y error plus index=3, y error minus index=2] {hpccg_data/hpccg_hpccg_100.dat} node[anchor=west,pos={12/24}, xshift=0, yshift=-5] {\footnotesize Hand-tuned};

\addplot table [x index=0, y index=1, y error plus index=3, y error minus index=2] {hpccg_data/hpccg_polly_100.dat} node[anchor=east,pos={1}, xshift=0, yshift=-5] {\footnotesize Polly};

\draw[very thick, pen colour={p_orange}, decorate,decoration={calligraphic brace, mirror, amplitude=2.3pt}] (12.2, 3.13) -- node[text=black, right=0.4em, fill=white] {\footnotesize 2.4x} (12.2, 7.63);

\addplot[black,dashed,domain=-2:26,samples = 2] {6.974597243} node[anchor=east, pos={3/4}, fill=white] {\footnotesize Serial};

% Plot 2
\nextgroupplot[
    xlabel={},
    ylabel={},
    error bars/y dir=both,
    error bars/y explicit,
    cycle list shift=1,
    title={Size $150^3$},
    xtick=data,
    clip=false,
    ymin=0,
    xmin=-2,
    xmax=26,
]
\addplot table [x index=0, y index=1, y error plus index=3, y error minus index=2] {hpccg_data/hpccg_dace_150.dat} node[anchor=east,pos={1}, xshift=0, yshift=-5] {\footnotesize DaCe};

\addplot table [x index=0, y index=1, y error plus index=3, y error minus index=2] {hpccg_data/hpccg_hpccg_150.dat} node[anchor=west,pos={16/24}, xshift=0, yshift=-5] {\footnotesize Hand-tuned};

\addplot table [x index=0, y index=1, y error plus index=3, y error minus index=2] {hpccg_data/hpccg_polly_150.dat} node[anchor=east,pos={1}, xshift=0, yshift=-5] {\footnotesize Polly};

\draw[very thick, pen colour={p_orange}, decorate,decoration={calligraphic brace, mirror, amplitude=3pt}] (24.5, 9.89) -- node[text=black, right=0.4em, fill=white] {\footnotesize 2.7x} (24.5, 27.13);

\addplot[black,dashed,domain=-2:26,samples = 2] {26.738836992} node[anchor=east, pos={3/4}, fill=white] {\footnotesize Serial};

% Plot 3
\nextgroupplot[
    xlabel={Threads},
    ylabel={Time [s]},
    error bars/y dir=both,
    error bars/y explicit,
    cycle list shift=1,
    title={Size $200^3$},
    xtick=data,
    clip=false,
    ymin=0,
    xmin=-2,
    xmax=26,
]
\addplot table [x index=0, y index=1, y error plus index=3, y error minus index=2] {hpccg_data/hpccg_dace_200.dat} node[anchor=east,pos={1}, xshift=0, yshift=-5] {\footnotesize DaCe};

\addplot table [x index=0, y index=1, y error plus index=3, y error minus index=2] {hpccg_data/hpccg_hpccg_200.dat} node[anchor=west,pos={19/24}, xshift=0, yshift=-5] {\footnotesize Hand-tuned};

\addplot table [x index=0, y index=1, y error plus index=3, y error minus index=2] {hpccg_data/hpccg_polly_200.dat} node[anchor=east,pos={1}, xshift=0, yshift=-5] {\footnotesize Polly};

\draw[very thick, pen colour={p_orange}, decorate,decoration={calligraphic brace, mirror, amplitude=3pt}] (24.5, 21.8) -- node[text=black, right=0.4em, fill=white] {\footnotesize 2.9x} (24.5, 63.24);

\addplot[black,dashed,domain=-2:26,samples = 2] {61.7865485705} node[anchor=east, pos={3/4}, fill=white] {\footnotesize Serial};

% Plot 4
\nextgroupplot[
    xlabel={Threads},
    ylabel={},
    error bars/y dir=both,
    error bars/y explicit,
    cycle list shift=1,
    title={Size $250^3$},
    xtick=data,
    clip=false,
    ymin=0,
    xmin=-2,
    xmax=26,
]
\addplot table [x index=0, y index=1, y error plus index=3, y error minus index=2] {hpccg_data/hpccg_dace_250.dat} node[anchor=east,pos={1}, xshift=0, yshift=-5] {\footnotesize DaCe};

\addplot table [x index=0, y index=1, y error plus index=3, y error minus index=2] {hpccg_data/hpccg_hpccg_250.dat} node[anchor=west,pos={21/24}, xshift=0, yshift=5] {\footnotesize Hand-tuned};

\addplot table [x index=0, y index=1, y error plus index=3, y error minus index=2] {hpccg_data/hpccg_polly_250.dat} node[anchor=east,pos={1}, xshift=0, yshift=-5] {\footnotesize Polly};

\draw[very thick, pen colour={p_orange}, decorate,decoration={calligraphic brace, mirror, amplitude=3pt}] (24.5, 43.82) -- node[text=black, right=0.4em, fill=white] {\footnotesize 2.9x} (24.5, 128.73);

\addplot[black,dashed,domain=-2:26,samples = 2] {118.94202787649999} node[anchor=east, pos={3/4}, fill=white] {\footnotesize Serial};

\end{groupplot}

\node (title) at ($(group c1r2.center)!0.5!(group c2r2.center)+(0,-2.8cm)$) {HPCCG with different problem sizes};

\end{tikzpicture}
    \caption{Runtime of Mantevo HPCCG benchmark with different compilers.}
\end{figure}

We can see that Polly was not able to find parallel opportunities as pointers are used to access the sparse matrix. Our approach was able to parallelize all five loops. The performance of our automatically parallelized code matches the hand-tuned version developers created across all experiments. At larger problem sizes, our approach even outperforms the reference implementation by up to 18\%.

\subsection{Lempel–Ziv–Oberhumer compression algorithm}
In Subsection~\ref{practical_results}, we presented results from a benchmark on the Lempel–Ziv–Oberhumer (LZO) compression algorithm. Although automatic parallelization wasn't achieved, there was still a modest enhancement in single-threaded performance. Achieving effective automatic parallelization would necessitate an algorithm redesign due to the presence of loop-carried dependencies. However, our method managed to achieve a 4\% runtime enhancement without requiring any algorithm rewriting. Given the analogous pointer movement patterns observed in most compression algorithms, we anticipate similar, if not superior, outcomes by extending our pipeline to encompass them as well.

\section{Related Work}
\paragraph{Pointer analysis in C}
Pointer analysis has been a topic of research for decades. One of the most impactful approaches, the "points-to" analysis, was proposed by Andersen et al.~\cite{Andersen2005ProgramAA} and developed and improved over the years~\cite{pt1,pt2,pt3}. Our approach is orthogonal to this method, and while more constrained in the patterns it detects, it provides a powerful benefit towards automatic parallelism detection. 

\vspace{-0.5em}
\paragraph{Parallelization of C programs.}
A large and increasing number of approaches exist that are geared toward finding  parallelism in C codes, with different degrees of automation.
Tools like DiscoPoP~\cite{li2015discopop} focus on identifying loops that could be parallelized using OpenMP constructs, while frameworks like Polly~\cite{polly} and Pluto~\cite{pluto} leverage the polyhedral representation of loops to automatically generate parallel code. Neither of these approaches can currently tackle codes using pointers as iterators, but we believe it should be possible for our approach to be leveraged in these workflows to increase their reach, as we have shown in this work with DaCe.

\vspace{-0.5em}
\paragraph{Parallelization of programs with pointers}
Utilizing the Distributed Dynamic Pascal (DDP) language~\cite{185487} facilitates the development of parallel programs involving pointers. The DDP language introduces explicit pointer access mechanisms to enable the parallelization of pointer-related operations. In contrast, our approach seamlessly integrates with existing C programs, obviating the necessity for transpilation processes.

\vspace{-0.5em}
\paragraph{Intermediate representations}
The LLVM IR~\cite{llvm} allows for many transformations and code optimizations. However, tracking pointer dependencies across scopes remains difficult. A new opportunity has appeared with the recent rise of MLIR~\cite{MLIR}, a framework that allows multiple IR dialects to co-exist and their different strengths to all contribute to overall code analyzability and performance. The MemRefType built-in dialect already enables to specify the shape of the underlying data. This includes the option to save the start of a data container even if the pointer is moved. MLIR would permit our approach to be implemented as a dialect, bringing it to the wider LLVM ecosystem.

\vspace{-0.5em}
\paragraph{Source code modification}
The alteration of source code within digital signal processing (DSP) programs facilitates the elimination of pointer references as outlined in the work by Franke et al. \cite{dsp}. While the approach bears some resemblance to our proposed method, it exhibits comparatively reduced flexibility. It exclusively encompasses static code modifications, devoid of the incorporation of the \textit{adjunct} concept, which enables the utilization of compile-time determined values exclusively. The primary aim remains the enhancement of computational efficiency, albeit without resorting to parallel processing techniques.

\vspace{-0.5em}
\paragraph{Pointers metadata}
Incorporating metadata into pointers is not a novel concept, as evidenced by previous studies~\cite{10.1145/3586038, 5601849}. In other scenarios, where the emphasis is on safety rather than performance, statically known mappings to pointers have been established. This facilitates the addition of runtime bounds checking, analogous to how we introduce runtime pointer movements through source code modifications.

\section{Conclusion}
In this work, we introduce a static transformation that separates the use of pointers as handles to data containers from their use as iterators. This improves code analyzability and data-centric compilers are shown to benefit from this transformation as eliminating  indirection exposes additional parallelization opportunities. Using our approach on the Mantevo HPCCG benchmark we were able to match the developer-optimized version and even surpass it by up to 6\%. On the OpenSSL PBKDF2 implementation, we were able to automatically parallelize it and obtain up to an 11x speedup.

\section*{Acknowledgments}
This project has received funding from EuroHPC-JU under grant agreement DEEP-SEA No 95560  and by the European Research Council (ERC) under the European Union’s Horizon 2020 program (grant agreement PSAP, No. 101002047). This work was partially supported by the ETH Future Computing Laboratory (EFCL), financed by a donation from Huawei Technologies.

\appendix
\section{Proof sketch of pointer assignment} \label{appendix_assignment}
We want to show that the assignment of pointers gives equivalent results after the application of the \textit{adjunct} transformation described in Figure~\ref{tab:transform}. To do so we introduce a new pointer $q$ with its $q_{adj}$. Let $n, m$ be arbitrary integers then we define $X, Y$ and $X', Y'$ as
\begin{align*}
X &= \left ( p := p + x_i | x_i \in \mathbb{Z}, 0 \leq i < n \right ) \\
Y &= \left ( v^i := v^i + x_i | x_i \in \mathbb{Z} \And v^i \in \{p, q\}, n \leq i < m \right ) \\
X' &= \left ( p_{adj} := p_{adj} + x_i | x_i \in \mathbb{Z}, 0 \leq i < n \right ) \\
Y' &= \left ( v^i_{adj} := v^i_{adj} + x_i | x_i \in \mathbb{Z} \And v^i_{adj} \in \{p_{adj}, q_{adj}\}, n \leq i < m \right )
\end{align*}
Using $;$ to concatenate commands we define $S$ with
$$S = (X; (q := p; Y))$$
Where it represents the execution of an arbitrary number of pointer movement statements followed by a pointer assignment and another arbitrary number of pointer movement statements. In the same way we define $S'$ as the \textit{adjunct} transformation of $S$.
$$S' = (X'; (q := p; (q_{adj} := p_{adj}; Y')))$$
Note that the $;$ operator is right associative. Then we define $Y'_q$ as
$$Y'_q = \sum^m_{i = n | v^i_{adj} = q_{adj}} x_i$$
i.e. the difference of value after executing $Y'$ on $q_{adj}$. Then $(Y'; q_{adj}) = q_{adj} + Y'_q$. Note that $X$ and $X'$ only act on $p$ and $p_{adj}$ respectively. This is to have a simpler proof but it can be easily shown that by defining $X$ and $X'$ in a similar way as $Y$ and $Y'$ the same result will be obtained.

For sake of a shorter syntax and more readable proof we omit the declaration of the pointers and \textit{adjunct} integer variables. I.e.
$$(p := malloc(...); (p_{adj} := 0; (q := malloc(...); (q_{adj} := 0; (S; q)))))$$
as it will be carried without modifications through all the proof steps. Then we can derive
\begin{align*}
    S; q &= X; (q := p; (Y; q)) \\
    \textit{(adjunct transformation)}      &= X; (q := p; (q + (Y'; q_{adj}))) \\
    \textit{(distributive)}             &= X; ((q := p; q) + (q := p; (Y'; q_{adj}))) \\
    \textit{(no effect statement)}      &= X; ((q := p; q) + (Y'; q_{adj})) \\
    \textit{(assignment)}               &= X; (p + (Y'; q_{adj})) \\
    \textit{(distributive)}             &= (X; p) + (X; (Y'; q_{adj})) \\
    \textit{(adjunct transformation)}      &= ((X'; p) + (X'; p_{adj})) + (X; (Y'; q_{adj})) \\
    \textit{(no effect statement)}      &= p + (X'; p_{adj}) + (X; (Y'; q_{adj})) \\
    \textit{(no effect statement)}      &= p + (X'; p_{adj}) + (Y'; q_{adj}) \\
    \textit{(sum expantion of adjunct)}    &= p + (X'; p_{adj}) + q_{adj} + Y'_q \\
    \textit{(adjunct value of 0)}          &= p + (X'; p_{adj}) + Y'_q \\
    \textit{(no effect statement)}      &= p + (X'; p_{adj}) + (X';Y'_q) \\
    \textit{(distributive)}             &= p + (X'; (p_{adj} + Y'_q)) \\
    \textit{(no effect statement)}      &= p + (X'; (q := p; (p_{adj} + Y'_q))) \\
    \textit{(assignment)}               &= p + (X'; (q := p; (q_{adj} := p_{adj}; q_{adj} + Y'_q)))) \\
    \textit{(}Y'_q\textit{ definition)}          &= p + (X'; (q := p; (q_{adj} := p_{adj}; (Y'; q_{adj})))) \\
    \textit{(S' definition)}            &= p + (S'; q_{adj}) \\
    \textit{(no effect statement)}      &= (X'; p) + (S'; q_{adj}) \\
    \textit{(assignment)}               &= (X'; (q := p; q)) + (S'; q_{adj}) \\
    \textit{(no effect statement)}      &= (X'; (q := p; (q_{adj} := p_{adj}; q))) + (S'; q_{adj}) \\
    \textit{(no effect statement)}      &= (X'; (q := p; (q_{adj} := p_{adj}; (Y'; q)))) + (S'; q_{adj}) \\
    \textit{(S' definition)}            &= (S';q) + (S'; q_{adj}) = q + (S'; q_{adj})
\end{align*}

Now we have proven the equivalence of the pointer assignment transformation stated in Figure~\ref{tab:transform}.

\bibliographystyle{ACM-Reference-Format}
\bibliography{references}

\end{document}